\pdfoutput=1

\documentclass[11pt,twoside,a4paper,cmspaper,final,collab]{cms-tdr}
\def\svnVersion{465343P}\def\cmsCernNoTag{CERN-EP-2017-308}\def\cmsCernDate{\today}\def\cmsMessage{Published in the European Physical Journal C as \href{http://dx.doi.org/10.1140/epjc/s10052-018-5950-6}{\doi{10.1140/epjc/s10052-018-5950-6.}}}

\begin{document}\cmsNoteHeader{HIN-16-025}

\hyphenation{had-ron-i-za-tion}
\hyphenation{cal-or-i-me-ter}
\hyphenation{de-vices}
\RCS$Revision: 461854 $
\RCS$HeadURL: svn+ssh://svn.cern.ch/reps/tdr2/papers/HIN-16-025/trunk/HIN-16-025.tex $
\RCS$Id: HIN-16-025.tex 461854 2018-05-27 20:36:13Z alverson $

\ifthenelse{\boolean{cms@external}}{\providecommand{\cmsLeft}{upper\xspace}}{\providecommand{\cmsLeft}{left\xspace}}
\ifthenelse{\boolean{cms@external}}{\providecommand{\cmsRight}{lower\xspace}}{\providecommand{\cmsRight}{right\xspace}}
\ifthenelse{\boolean{cms@external}}{\providecommand{\cmsLLeft}{Upper\xspace}}{\providecommand{\cmsLLeft}{Left\xspace}}
\ifthenelse{\boolean{cms@external}}{\providecommand{\cmsRRight}{Lower\xspace}}{\providecommand{\cmsRRight}{Right\xspace}}

\newcommand{\sqrts}{\ensuremath{\sqrt{s}}\xspace}
\newcommand{\mumu}{\ensuremath{\PGmp\PGmm}\xspace}
\newcommand{\psiP}{\ensuremath{\psi\text{(2S)}}\xspace}
\renewcommand{\PgU}{\ensuremath{\Upsilon}\xspace}
\renewcommand{\PgUa}{\ensuremath{\Upsilon\text{(1S)}}\xspace}
\newcommand {\npart}{\ensuremath{N_{\text{part}}}\xspace}
\newcommand{\raa}{\ensuremath{R_{\text{AA}}}\xspace}
\newcommand{\taa}{\ensuremath{T_{\text{AA}}}\xspace}
\newcommand{\nmb}{\ensuremath{N_{\text{MB}}}\xspace}
\newcommand{\pp}{{\ensuremath{\Pp\Pp}}\xspace}
\newcommand{\PbPb}{\ensuremath{\text{PbPb}}\xspace}
\newcommand{\AuAu}{\ensuremath{\text{AuAu}}\xspace}
\newcommand{\sqrtsnn}{\ensuremath{\sqrt{\smash[b]{s_{_{\text{NN}}}}}}\xspace}
\newcommand{\Lxyz}{\ensuremath{L_{xyz}}\xspace}
\newcommand{\ctauxyz}{\ensuremath{\ell_{\JPsi}}\xspace}

\cmsNoteHeader{HIN-16-025}
\title{Measurement of prompt and nonprompt charmonium suppression in \PbPb collisions at 5.02\TeV}

\date{\today}

\abstract{
The nuclear modification factors of \JPsi and \psiP mesons are measured in \PbPb collisions at a centre-of-mass energy per nucleon pair of $\sqrtsnn = 5.02\TeV$. The analysis is based on \PbPb and \pp data samples collected by CMS at the LHC in 2015, corresponding to integrated luminosities of 464\mubinv and 28\pbinv, respectively. The measurements are performed in the dimuon rapidity range of $\abs{y} < 2.4$ as a function of centrality, rapidity, and transverse momentum (\pt{}) from $\pt=3$\GeVc in the most forward region and up to 50\GeVc. Both prompt and nonprompt (coming from b hadron decays) \JPsi mesons are observed to be increasingly suppressed with centrality, with a magnitude similar to the one observed at $\sqrtsnn= 2.76\TeV$ for the two \JPsi meson components. No dependence on rapidity is observed for either prompt or nonprompt \JPsi mesons. An indication of a lower prompt \JPsi meson suppression at $\pt > 25$\GeVc is seen with respect to that observed at intermediate \pt. The prompt \psiP meson yield is found to be more suppressed than that of the prompt \JPsi mesons in the entire \pt range.
}

\hypersetup{%
pdfauthor={CMS Collaboration},%
pdftitle={Measurement of prompt and nonprompt charmonium suppression in PbPb collisions at 5.02 TeV},%
pdfsubject={CMS},%
pdfkeywords={CMS, physics, heavy ions, quarkonia, J/psi, psi(2S), PbPb, R[AA], 5 TeV}}

\maketitle

\section{Introduction}

Quarkonium production in heavy ion collisions has a rich history. In their original article~\cite{Matsui:1986dk}, Matsui and Satz proposed that Debye color screening of the heavy-quark potential in a hot medium prevents the production of \JPsi mesons (and this applies also to other heavy-quark bound states such as \psiP, and \PgUa\ mesons \cite{Digal:2001ue}). Consequently, the suppression of quarkonium yields in heavy ion collisions, relative to those in \pp collisions, has long been considered to be a sensitive probe of deconfinement and quark-gluon plasma formation. The \JPsi meson suppression observed in \PbPb collisions at the CERN SPS~\cite{Alessandro:2004ap} and \AuAu collisions at the BNL RHIC~\cite{Adare:2006ns} is compatible with this picture. Similarly, the disappearance of \PgU\ resonances in \PbPb collisions at the CERN LHC~\cite{Chatrchyan:2012lxa,Abelev:2014nua} is consistent with the Debye screening scenario.

When produced abundantly in a single heavy ion collision, uncorrelated heavy quarks may combine to form quarkonia states in the medium~\cite{BraunMunzinger:2000px,Thews:2000rj}. This additional source of quarkonium, commonly referred to as \emph{recombination}, would enhance its production in heavy ion collisions, in contradistinction with the Debye screening scenario. Signs of this effect can be seen in the recent results from the ALICE Collaboration at the LHC~\cite{Abelev:2012rv,Adam:2016rdg}, which measured a weaker \JPsi meson suppression than at RHIC~\cite{Adare:2006ns,Adare:2011yf}, despite the higher medium energy density. Note that recombination is only expected to affect charmonium production at low transverse momenta ($p_\text{T}$), typically for values smaller than the charmonium mass ($\pt \lesssim m_{\psi} \, c$), where the number of charm quarks initially produced in the collision is the largest \cite{Thews:2000rj}.

At large \pt, other mechanisms may contribute to charmonium suppression. Until recently, no quarkonium results were available at high $\pt$, because of kinematic constraints at the SPS and too low counting rates at RHIC. At the LHC, a strong \JPsi suppression has been measured up to $\pt = 30$\GeVc by the CMS Collaboration~\cite{Khachatryan:2016ypw} in \PbPb collisions at a centre-of-mass energy per nucleon pair of $\sqrtsnn = 2.76\TeV$. Results at $5.02\TeV$ have also been reported, up to $\pt = 10$\GeVc, by the ALICE Collaboration~\cite{Adam:2016rdg}.
According to Refs.~\cite{Strickland:2011mw,Du:2015wha}, quarkonium suppression by Debye screening may occur even at high $\pt$. At the same time, when $\pt \gg m_{\psi} \, c$, heavy quarkonium is likely to be produced by parton fragmentation, hence it should rather be sensitive to the parton energy loss in the quark-gluon plasma. The similarity of \JPsi meson suppression with the quenching of jets, light hadrons, and D mesons supports this picture~\cite{Spousta:2016agr,Arleo:2017ntr,Khachatryan:2016ypw}.

At the LHC, the inclusive \JPsi meson yield also contains a significant \textit{nonprompt} contribution coming from \PQb hadron decays~\cite{Aaij:2011jh,Khachatryan:2010yr,Aad:2011sp}. The nonprompt \JPsi component should reflect medium effects on \PQb hadron production in heavy ion collisions, such as b quark energy loss. Measuring both prompt and nonprompt \JPsi meson production in \PbPb collisions thus offers the opportunity to study both hidden charm and open beauty production in the same data sample.

In this paper we report on a new measurement of the prompt and nonprompt \JPsi and \psiP nuclear modification factors (\raa) using \PbPb data, collected at the end of 2015 with the CMS experiment at $\sqrtsnn = 5.02\TeV$. The analysis is performed via the dimuon decay channel. The results are compared to those obtained at 2.76\TeV~\cite{Khachatryan:2016ypw}. The larger integrated luminosities allow for more precise and more differential measurements of \raa, as functions of centrality, rapidity ($y$), and \pt up to 50\GeVc.

\section{The CMS detector}

The central feature of the CMS apparatus is a superconducting solenoid of 6\unit{m} internal diameter, providing a magnetic field of 3.8\unit{T}. Within the solenoid volume are a silicon pixel and strip tracker, a lead tungstate crystal electromagnetic calorimeter, and a brass and scintillator hadron calorimeter, each composed of a barrel and two endcap sections. Forward calorimeters extend the coverage provided by the barrel and endcap detectors. Muons are measured in the pseudorapidity range $\abs{\eta} < 2.4$ in gas-ionisation detectors embedded in the steel flux-return yoke outside the solenoid, with detection planes made using three technologies: drift tubes, cathode strip chambers, and resistive-plate chambers. The hadron forward (HF) calorimeters use steel as an absorber and quartz fibres as the sensitive material. The two HF calorimeters are located 11.2\unit{m} from the interaction region, one on each side, and together they provide coverage in the range $2.9 < \abs{\eta} < 5.2$. They also serve as luminosity monitors. Two beam pick-up timing detectors are located at 175\unit{m} on both sides of the interaction point, and provide information about the timing structure of the LHC beam.
Events of interest are selected using a two-tiered trigger system~\cite{Khachatryan:2016bia}. The first level (L1), composed of custom hardware processors, uses information from the calorimeters and muon detectors to select events. The second level, known as the high-level trigger (HLT), consists of a farm of processors running a version of the full event reconstruction software optimised for fast processing.
A more detailed description of the CMS detector, together with a definition of the coordinate system used and the relevant kinematic variables, can be found in Ref.~\cite{Chatrchyan:2008zzk}.

For \pp data the vertices are reconstructed with a deterministic annealing vertex fitting algorithm using all of the fully reconstructed tracks~\cite{Chatrchyan:2014aa}. The physics objects used to determine the primary vertex are defined based on a jet finding algorithm~\cite{Cacciari:2008gp,Cacciari:2011ma} applied to all charged tracks associated with the vertex, plus the corresponding associated missing transverse momentum. The reconstructed vertex with the largest value of summed physics object $\pt^2$ is taken to be the primary \pp interaction vertex. In the case of \PbPb data, a single primary vertex is reconstructed using a gap clustering algorithm~\cite{Chatrchyan:2014aa}, using pixel tracks only.

 \section{Data selection}

 \subsection{Event selection}

Hadronic collisions are selected offline using information from the HF calorimeters. In order to select \PbPb collisions, at least three towers with energy deposits above 3\GeV are required in each of the HF calorimeters, both at forward and backward rapidities.
A primary vertex reconstructed with at least two tracks is also required. In addition, a filter on the compatibility of the silicon pixel cluster width and the vertex position is applied~\cite{Khachatryan:2010aa}. The combined efficiency for this event selection, including the remaining non-hadronic contamination, is $(99 \pm 2)$\%. Values higher than 100\% are possible, reflecting the possible presence of ultra-peripheral (i.e. non-hadronic) collisions in the selected event sample.

The \PbPb sample is divided into bins of collision centrality, which is a measure of the degree of overlap of the colliding nuclei and is related to the number of participating nucleons (\npart). Centrality is defined as the percentile of the inelastic hadronic cross section corresponding to a HF energy deposit above a certain threshold~\cite{Chatrchyan:2011pb}. The most central (highest HF energy deposit) and most peripheral (lowest HF energy deposit) centrality bins used in
the analysis are 0--5\% and 70--100\% respectively. Variables related to the centrality, such as \npart and the nuclear overlap function (\taa)~\cite{Khachatryan:2016odn}, are estimated using a Glauber model simulation described in Ref.~\cite{Miller:2007ri}.

The \pp and \PbPb data sets correspond to integrated luminosities of 28.0\pbinv and 464\mubinv, respectively.
Both \JPsi and \psiP mesons are reconstructed using their dimuon decay channel.
The dimuon events were selected online by the L1 trigger system, requiring two tracks in the muon detectors with no explicit momentum threshold, in coincidence with a bunch crossing identified by beam pick-up timing detectors.
No additional selection was applied by the HLT.
Because of the high rate of the most central dimuon events, a prescale was applied at the HLT level during part of the PbPb data taking: as a consequence only 79\% of all the dimuon events were recorded, resulting in an effective luminosity of 368\mubinv.
For peripheral events we were able to sample the entire integrated luminosity of 464\mubinv. This was done by adding an additional requirement that events be in the centrality range of 30--100\% to the dimuon trigger.
The prescaled data sample is used for the results integrated over centrality and those in the centrality range 0--30\%, while for the results in the 30--100\% range the data sample with 464\mubinv was used instead.
The results reported in this paper are unaffected by the small number of extra collisions potentially present in the collected events: the mean of the Poisson distribution of the number of collisions per bunch crossing (pileup), averaged over the full data sample, is approximately 0.9 for the \pp data and less than 0.01 for \PbPb collisions.

Simulated events are used to tune the muon selection criteria and the signal fitting parameters, as well as for acceptance and efficiency studies. These samples, produced using \PYTHIA 8.212~\cite{Sjostrand2015}, and decaying the \PQb hadrons with \textsc{evtgen} 1.3.0~\cite{Lange:2001uf}, are embedded in a realistic \PbPb background event generated with \textsc{hydjet} 1.9~\cite{Lokhtin:2005px} and propagated through the CMS detector with \GEANTfour~\cite{Agostinelli:2002hh}. The prompt \JPsi is simulated unpolarised, a scenario in good agreement with pp measurements~\cite{Abelev:2011md,Chatrchyan:2013cla,Aaij:2013nlm}. For nonprompt \JPsi, the polarisation is the one predicted by \textsc{evtgen}, roughly $\lambda_{\theta} = 0.4$. The resulting events are processed through the trigger emulation and the event reconstruction sequences. The assumptions made on the quarkonium polarisation affect the computation of the acceptance. Quantitative estimates of the possible effect evaluated for several polarisation scenarios can be found in Refs. \cite{Chatrchyan:2012np,Sirunyan:2018aa}. While there are no measurements on quarkonium polarisations in \PbPb collisions, a study in \pp collisions as a function of the event activity \cite{Khachatryan:2016aa} has not revealed significant changes. Therefore the effects of the \JPsi polarisation on the acceptance are not considered as systematic uncertainties.

 \subsection{Muon selection}

The muon reconstruction algorithm starts by finding tracks in the muon detectors, which are then fitted together with tracks reconstructed in the silicon tracker. Kinematic selections are imposed to single muons so that their combined trigger, reconstruction and identification efficiency stays above 10\%. These selections are: $\pt^{\PGm}>3.50\GeVc$ for $\abs{\eta^{\PGm}}<1.2$ and $\pt^{\PGm}>1.89\GeVc$ for $2.1<\abs{\eta^{\PGm}}<2.4$, linearly interpolated in the intermediate $\abs{\eta^{\PGm}}$ region.
The muons are required to match the ones selected by the dimuon trigger, and \emph{soft} muon selection criteria are applied to \emph{global muons} (i.e. muons reconstructed using the combined information of the tracker and muon detectors), as defined in Ref.~\cite{Chatrchyan:2012xi}. Matching muons to tracks measured in the silicon tracker results in a relative \pt resolution for muons between 1 and 2\% for a typical muon in this analysis~\cite{Chatrchyan:2012xi}.
In order to remove cosmic and in-flight decay muons, the transverse and longitudinal distances of approach to the measured vertex of the muons entering in the analysis are required to be less than 0.3 and 20~cm, respectively. The probability that the two muon tracks originate from a common vertex is required to be larger than 1\%, lowering the background from \PQb and \PQc hadron semileptonic decays.

 \section{Signal extraction}

Because of the long lifetime of \PQb hadrons compared to that of \JPsi mesons, the separation of the prompt and nonprompt \JPsi components relies on the measurement of a secondary $\PGmp\PGmm$ vertex displaced from the primary collision vertex. The \JPsi mesons originating from the decay of \PQb hadrons can be resolved using the pseudo-proper decay length~\cite{Buskulic:1993aa} $\ctauxyz =  \Lxyz \: m_{\JPsi} \: c / |p_{\mu\mu}|$, where $\Lxyz$ is the distance between the primary and dimuon vertices, $m_{\JPsi}$ is the Particle Data Group \cite{pdg} world average value of the \JPsi meson mass (assumed for all dimuon candidates), and $p_{\mu\mu}$ is the dimuon momentum. Note that due to resolution effects and background dimuons the pseudo-proper decay length can take negative values. To measure the fraction of \JPsi mesons coming from \PQb hadron decays (the so-called nonprompt fraction), the invariant mass spectrum of $
\PGmp\PGmm$ pairs and their \ctauxyz distribution are fitted using a two-dimensional (2D) extended unbinned maximum-likelihood fit. In order to obtain the parameters of the different components of the 2D probability density function (PDF), the invariant mass and the \ctauxyz distributions are fitted sequentially prior to the final 2D fits, as explained below.
These fits are performed for each \pt, rapidity and centrality bin of the analysis, and separately in \pp and \PbPb collisions.

\begin{figure}[htbp]
  \centering
    \includegraphics[width=0.48\textwidth]{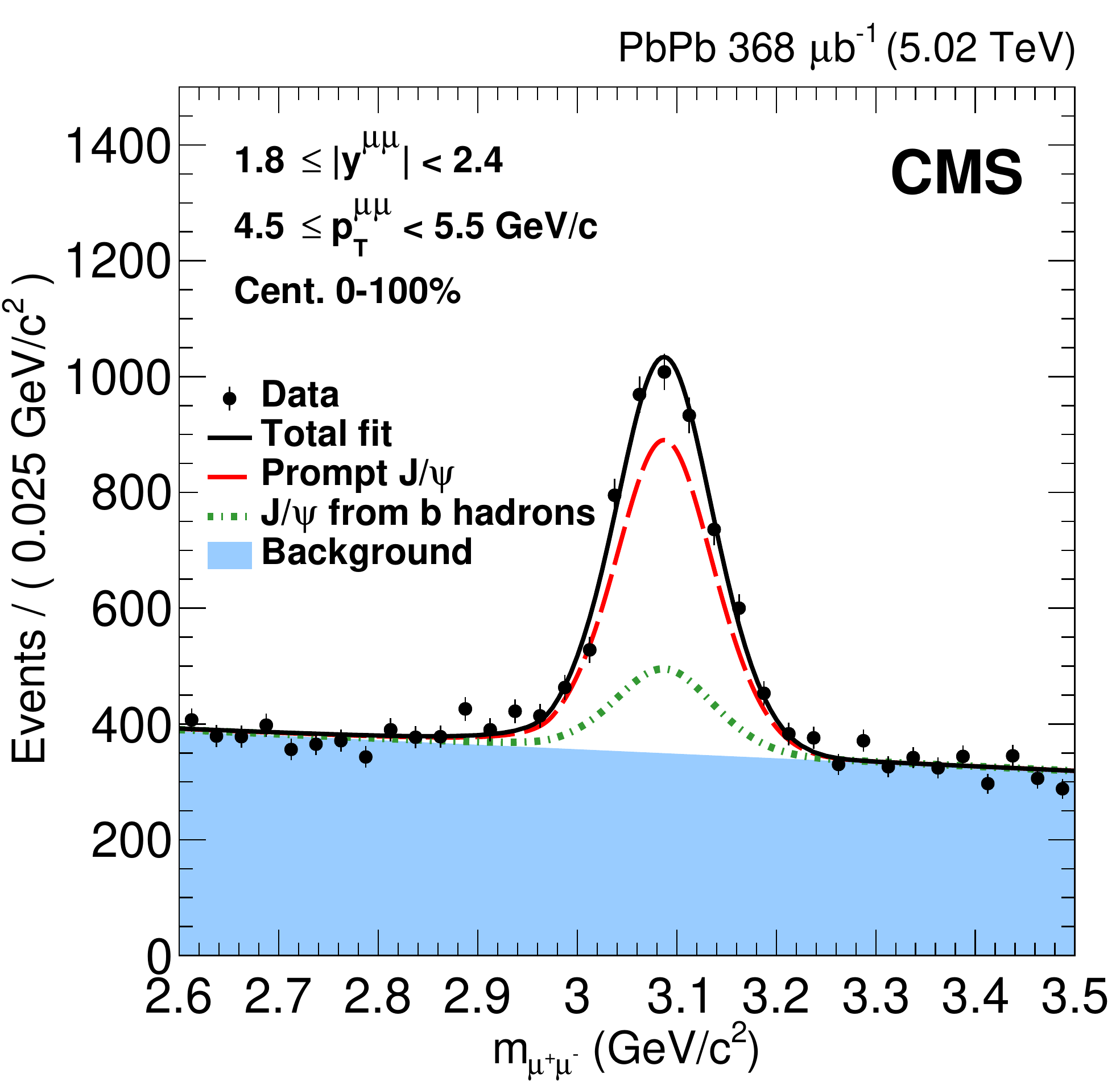}
        \includegraphics[width=0.48\textwidth]{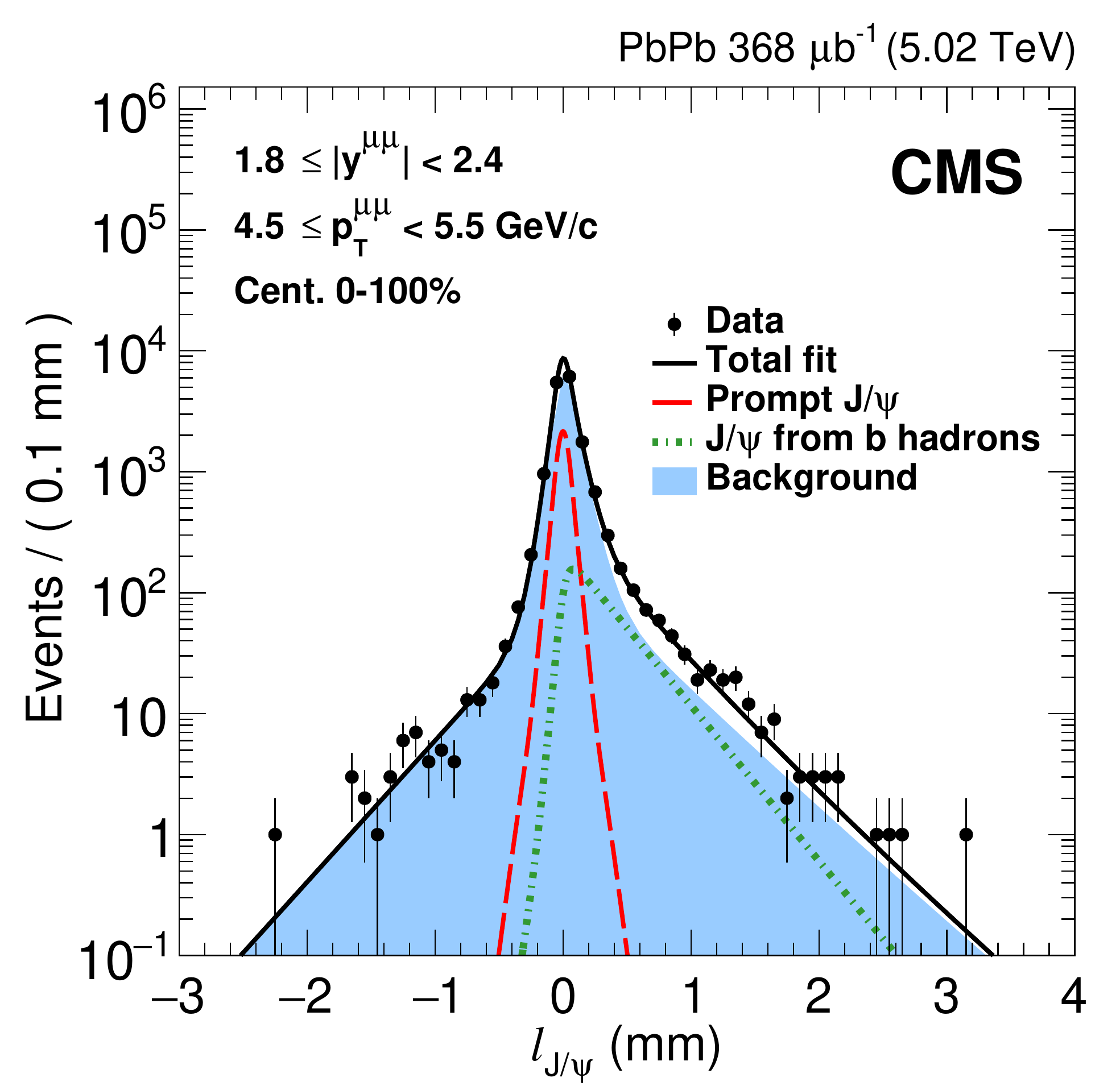}
    \caption{Invariant mass spectrum of \mumu pairs (\cmsLeft) and pseudo-proper decay length distribution (\cmsRight) in \PbPb collisions for $1.8<\abs{y}<2.4$, $4.5<\pt<5.5\GeVc$, for all centralities. The result of the fit described in the text is also shown.
      }
    \label{fig:2Dfits}
\end{figure}

The sum of two Crystal Ball functions~\cite{SLAC-R-236}, with different widths but common mean and tail parameters, is used to extract the nominal yield values from the \pp and \PbPb invariant mass distributions. The tail parameters, as well as the ratio of widths in the \PbPb case, are fixed to the values obtained from simulation. The background is described by a polynomial function of order $N$, where $N$ is the lowest value that provides a good description of the data, and is determined by performing a log-likelihood ratio test between polynomials of different orders, in each analysis bin, while keeping the tail and width ratio parameters fixed. The order of the polynomial is chosen in such a way that increasing the order does not significantly improve the quality of the fit. The typical order of the polynomial is 1 for most of the analysis bins. The invariant mass signal and background parameters are obtained in an initial fit of the invariant mass distribution only and then fixed on the 2D fits of mass and \ctauxyz distributions, while the number of extracted \JPsi mesons and background dimuons are left as free parameters.

The prompt, nonprompt, and background components of the \ctauxyz distributions are parameterised using collision data and Monte Carlo (MC) simulated events, and the signal and background contributions unfolded with the ${}_{s}\mathcal{P}lot$ technique \cite{sPlot}. In the context of this analysis, this technique uses the invariant mass signal and background PDFs to discriminate signal from background in the \ctauxyz distribution. The \ctauxyz per-event uncertainty distributions of signal and background, provided by the reconstruction algorithm of primary and secondary vertices, are extracted from data and used as templates.
The \ctauxyz resolution is also obtained from the data by fitting the distribution of events with $\ctauxyz < 0$ with a combination of three Gaussian functions. The resolution varies event-by-event, so the per-event uncertainty is used as the width of the Gaussian function that describes the core. To take into account the difference on the per-event uncertainty distributions of signal and background dimuons, the resolution PDF is multiplied by the per-event uncertainty distribution of signal and background dimuons separately. All the resolution parameters are fixed in the 2D fits.
The \PQb hadron decay length is allowed to float freely in the fit, and it is initialised to the value extracted by fitting the \ctauxyz distribution of nonprompt \JPsi mesons from a MC sample with an exponential decay function, at generator level. The \ctauxyz distribution of background dimuons is obtained from fits to the data, using an empirical combination of exponential functions. The parameters of the \ctauxyz background distribution are also fixed in the 2D fits. Finally, the number of extracted \JPsi mesons, the number of background dimuons and the nonprompt fraction are extracted from the 2D fits. An example of a 2D fit of the invariant mass and pseudo-proper decay length for the \PbPb data is shown in Fig.~\ref{fig:2Dfits} for a representative analysis bin.

\section{Acceptance and efficiency corrections}

Correction factors are applied to all results to account for detector acceptance,
 trigger, reconstruction, and selection efficiencies of the $\PGmp\PGmm$
pairs.
The corrections are derived from prompt and nonprompt \JPsi meson MC samples in \pp and \PbPb, and are evaluated in the same bins of \pt, centrality, and rapidity used in the \raa and cross section analyses.
The prompt and nonprompt \JPsi meson \pt distributions in bins of rapidity in MC samples are compared
to those in data, and the ratios of data over MC are used to weight the MC \JPsi distributions to describe the data better.
This weighting accounts for possible mis-modelling of \JPsi kinematics in MC.
The acceptance in a given analysis bin is defined as the fraction of generated \JPsi mesons in that bin which decay into two muons entering the kinematic limits defined above, and reflects the geometrical coverage of the CMS detector. The value of the acceptance correction ranges from 4 to 70\%, depending on the dimuon \pt, both for prompt and nonprompt \JPsi mesons in \pp and \PbPb collisions.
The efficiency in a given analysis bin is defined as the ratio of the number of reconstructed \JPsi mesons in which both muons pass the analysis selection and the number of generated \JPsi mesons in which both muons pass the analysis selection.
The efficiency correction depends on the dimuon \pt, rapidity and event centrality, and ranges from 20 to 75\% (15 to 75\%) for prompt (nonprompt) \JPsi mesons in \PbPb data, and from 40 to 85\% for both prompt and nonprompt \JPsi mesons in \pp data. The efficiency is lower at low than at high \pt, and it decreases from mid to forward rapidity; it is also lower for central than peripheral events.
The individual components of the efficiency (tracking reconstruction,
standalone muon reconstruction, global muon fit, muon identification and
selection, and triggering) are also measured using single muons from \JPsi meson
decays in both simulated and collision data, using the \emph{tag-and-probe} (T$\&$P) technique~\cite{Khachatryan:2011aa,Chatrchyan:2012np}. The values obtained from data and simulation are seen to differ only for the muon trigger efficiency and the ratio of the data over simulated efficiencies is used as a correction factor for the efficiency. The correction factor for dimuons is at most 1.35 (1.38) for the \pp (\PbPb) efficiency in the $3 < \pt < 4.5$\GeVc and forward rapidity bin, but the \pt and rapidity integrated value of the correction is about 1.03. The other T$\&$P efficiency components are compatible, hence only used as a cross-check, as well as to estimate systematic uncertainties.

\section{Systematic uncertainties}

The systematic uncertainties in these measurements arise from the invariant mass signal and background fitting model assumptions, the parameterisation of the \ctauxyz distribution, the acceptance and efficiency computation, and sample normalisation (integrated luminosity in \pp data, counting of the equivalent number of minimum bias events in \PbPb, and nuclear overlap function). These systematic uncertainties are derived separately for \pp and \PbPb results, and the total systematic uncertainty is computed as the quadratic sum of the partial terms.

The systematic uncertainty due to each component of the 2D fits is estimated from the difference between the nominal value and the result obtained with the variations of the different components mentioned below, in the extracted number of prompt and nonprompt \JPsi mesons, or nonprompt fraction separately. In the following, the typical uncertainty is given for the observable on which each source has the biggest impact.

In order to determine the uncertainty associated with the invariant mass fitting procedure, the signal and background PDFs are independently varied, in each analysis bin. For the uncertainty in the signal, the parameters that were fixed in the nominal fits are left free with a certain constraint. The constraint for each parameter is determined from fits to the data, by leaving only one of the parameters free, and it is chosen as the root mean square of the variations over the different analysis bins.
A different signal shape is also used: a Crystal Ball function plus a Gaussian function, with the CB tail parameters, as well as the ratio of widths in the PbPb case, again fixed from MC. The dominant uncertainty comes from the variation of the signal shape, yielding values for the number of extracted nonprompt \JPsi mesons ranging from 0.1 to 2.9\%  (0.3 to 5.5\%) in \pp (\PbPb) data.
For the background model, the following changes are considered, while keeping the nominal signal shape. First, the log-likelihood ratio tests are done again with two variations of the threshold used to choose the order of the polynomial function in each analysis bin. Also the fitted mass range is varied. Finally, an exponential of a polynomial function is also used. The dominant uncertainty in the background model arises from the assumed shape (invariant mass range) in \pp (\PbPb) data. The corresponding uncertainty ranges from 0.1 to 2.1\% (0.1 to 2.8\%).
The maximum difference of each of these variations, in each analysis bin and separately for the signal and the background, is taken as an independent systematic uncertainty.

For the \ctauxyz distribution fitting procedure, four independent variations of the different components entering in the 2D fits are considered.
For the \ctauxyz uncertainty distribution, instead of using the distributions corresponding to signal and background, the total distribution is assumed. The contribution to the systematic uncertainty in the number of extracted nonprompt \JPsi mesons ranges from 0.3 to 2\% (0.3 to 9.5\%) in \pp (\PbPb) data.
The \ctauxyz resolution obtained from prompt \JPsi meson MC is used instead of that evaluated from data. The corresponding uncertainty in the nonprompt fraction ranges from 1 to 5\% (1 to 11\%) in \pp (\PbPb) data.
A nonprompt \JPsi meson MC template replaces the exponential decay function for the \PQb hadron decay length. In this case, the contribution of this source to the systematic uncertainty in the nonprompt \JPsi  yield ranges from 0.2 to 8\% (0.2 to 20\%) in \pp (\PbPb) data.
A template of the \ctauxyz distribution of background dimuons obtained from the data is used to describe the background, instead of the empirical combination of exponential functions. This variation has an impact on the nonprompt \JPsi yield ranging from 0.1 to 1.3\% (0.2 to 22\%) in \pp (\PbPb) data.
Therefore the dominant sources of uncertainty in the \ctauxyz fitting are the background parameterisation and the MC template for the nonprompt signal. They have an important impact on the nonprompt \JPsi meson yield, especially at the lowest \pt reached in this analysis for the most central events in \PbPb collisions. The reason for this is that the background dimuons largely dominate over the nonprompt \JPsi signal.

The uncertainties in the acceptance and efficiency determination are evaluated with MC studies considering a broad range of \pt and angular spectra compatible with the pp and \PbPb data within their uncertainties. These variations yield an uncertainty about 0.2\% ($<$1.7\%) in \pp (\PbPb) collisions, both for prompt and nonprompt \JPsi acceptance and efficiency. The statistical uncertainty of the weighting of the MC distributions, reflecting the impact of the limited knowledge on the kinematic distribution of \JPsi mesons on the acceptance and efficiency corrections, is used as systematic uncertainty.  This uncertainty is at most 6\% (11\%) in \pp (\PbPb) collisions at the largest \pt but it usually ranges from 1 to 3\% in both collision systems.
In addition, the systematic uncertainties in the T\&P correction factors, arising from the limited data sample available and from the procedure itself, are taken into account, covering all parts of the muon efficiency: inner tracking and muon reconstruction, identification, and triggering. The dominant uncertainty in the T\&P correction factors arises from muon reconstruction and ranges from 2 to 10\% for both collision systems.

The global uncertainty in the pp luminosity measurement is 2.3\%~\cite{CMS-PAS-LUM-16-001}. The number of minimum bias events corresponding to our dimuon sample in PbPb ($N_{\mathrm{MB}}$) comes from a simple event counting in the events selected by the Minimum Bias triggers, taking into account the trigger prescale. The corresponding uncertainty arises from the inefficiency of trigger and event selection, and is estimated to be 2\%.
Finally, the uncertainty in the \taa is estimated by varying the Glauber model parameters within their uncertainty and taking into account the uncertainty on the trigger and event selection efficiency, and ranges from 3 to 16\% from the most central to the most peripheral events used in this analysis.

 \section{Results}

In this section, the results obtained for nonprompt \JPsi fractions, prompt and nonprompt \JPsi cross sections for each collision system, and nuclear modification factors \raa are presented and discussed. In addition, a derivation of the \psiP \raa is also presented and discussed.
For all results plotted versus \pt or $\abs{y}$, the abscissae of the points correspond to the centre of the respective bin, and the horizontal error bars reflect the width of the bin. The lower \pt thresholds in the different rapidity intervals reflect the detector acceptance. In the range $1.8 < \abs{y} < 2.4$ \JPsi are measured down to 3\GeVc, while for the bins with $\abs{y} < 1.8$ they are measured down to 6.5\GeVc.
When plotted as a function of centrality, the abscissae are the average \npart values for minimum bias events within each centrality bin.
The weighted average \npart values (weighted for the number of binary nucleon-nucleon collisions) correspond in most cases to the average \npart values for minimum bias events, with the exception of the most peripheral bin (50--100\%) where \npart changes from 22 to 43.
The centrality binning used is 0--5--10--15--20--25--30--35--40--45--50--60--70--100\% for the results in $\abs{y}<2.4$, and  0--10--20--30--40--50--100\% for the results differential in rapidity.

\subsection{Nonprompt \texorpdfstring{\JPsi}{J/psi} meson fractions}

\begin{figure}[hbtp]
  \centering
\includegraphics[width=0.48\textwidth]{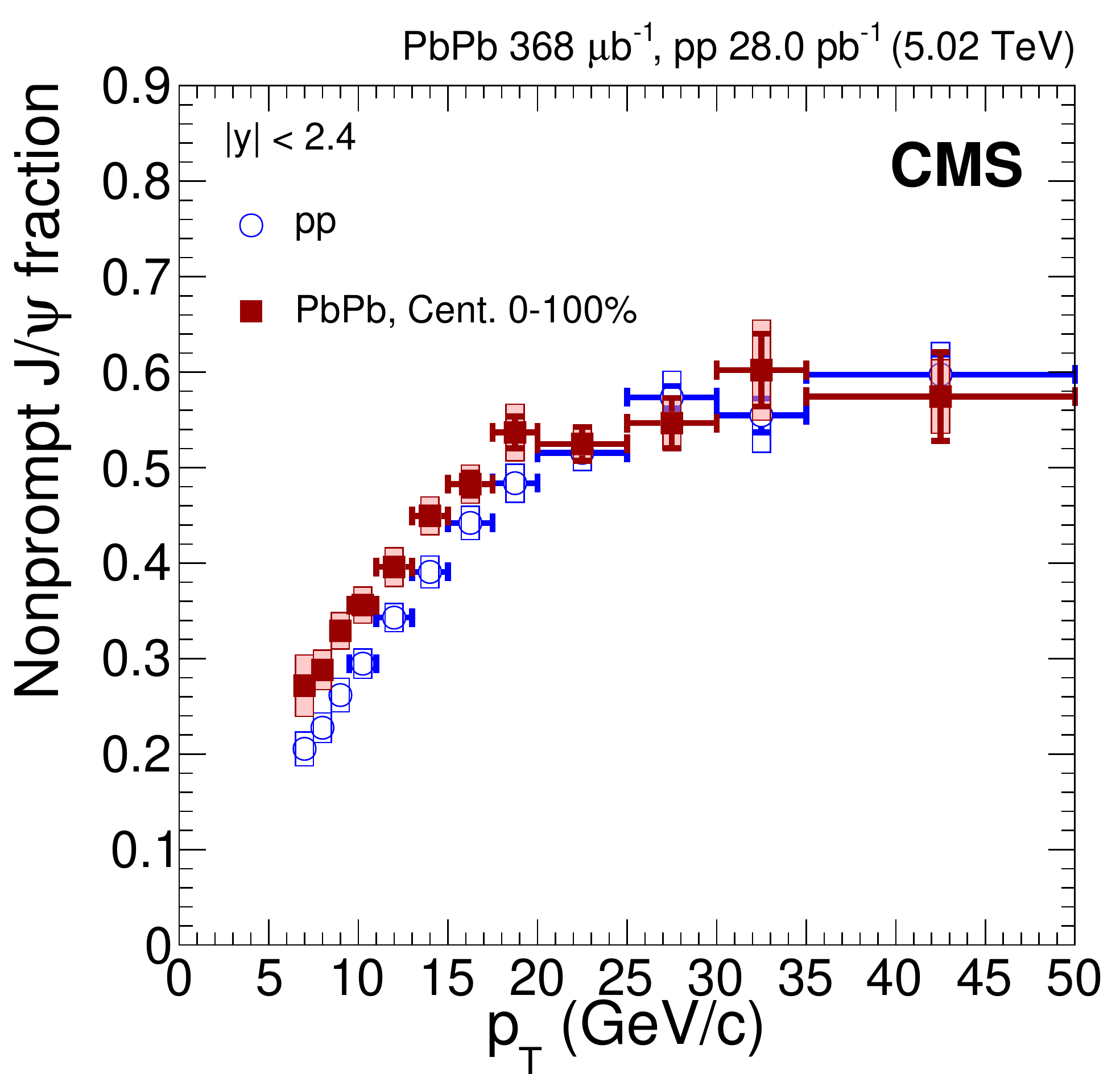}
\includegraphics[width=0.48\textwidth]{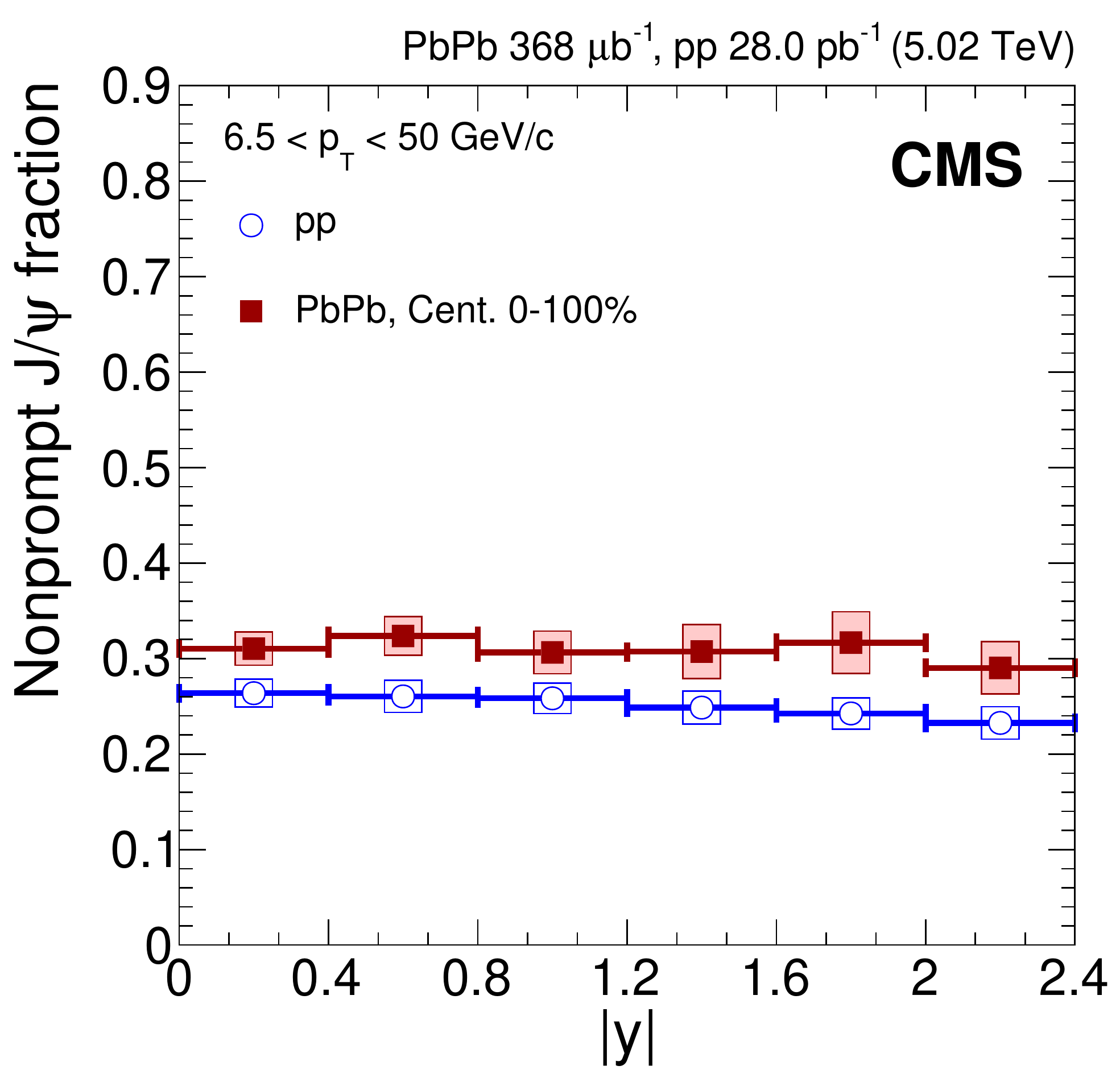}
  \caption{Fraction of \JPsi mesons coming from the decay of b hadrons, i.e. nonprompt \JPsi meson fraction, as a function of dimuon \pt (\cmsLeft) and rapidity (\cmsRight) for \pp and \PbPb collisions, for all centralities. The bars (boxes) represent statistical (systematic) point-by-point uncertainties.}
    \label{fig:bFracspp}
\end{figure}

The nonprompt \JPsi meson fraction is defined as the proportion of measured \JPsi mesons coming from b hadron decays, corrected for acceptance and efficiency. It is presented in Fig.~\ref{fig:bFracspp} for \pp and \PbPb collisions, as a function of \pt and rapidity, in the full $\abs{y}<2.4$ and $6.5<\pt<50\GeVc$ range. No significant rapidity dependence is observed, while there is a strong \pt dependence, from about 20\% at low \pt to 60\% at high \pt, reflecting the different \pt distributions of prompt and nonprompt \JPsi mesons, which highlights the necessity of separating the two contributions.

 \subsection{Prompt and nonprompt \texorpdfstring{\JPsi}{J/psi} meson cross sections in \pp and \PbPb collisions}

The measurements of the prompt and nonprompt \JPsi cross sections can help to test the existing theoretical models of both quarkonium production and b hadron production.
The cross sections are computed from the corrected yields,
\begin{equation}
 \frac{\rd^2 N}{\rd\pt\,\rd{}y} = \frac{1}{\Delta \pt\, \Delta y} \, \frac{N_{\JPsi}}{\mathcal{A} \, \epsilon},
\end{equation}
where $N_{\JPsi}$ is the number of prompt or nonprompt \JPsi mesons, $\mathcal{A}$ is the acceptance, $\epsilon$ is the efficiency, and $\Delta \pt$ and $\Delta y$ are the \pt and rapidity bin widths, respectively. To put the \pp and \PbPb data on a comparable scale, the corrected yields are normalised by the measured integrated luminosity for pp collisions ($\sigma = N / \mathcal{L}$), and by the product of the number of corresponding minimum bias events and the centrality-integrated nuclear overlap value  for \PbPb collisions ($N/(N_{\mathrm{MB}}\taa)$). Global  uncertainties (common to all measurements) arise from these normalisation factors and account for the integrated luminosity uncertainty in \pp collisions ($\pm$2.3\%) and the \nmb and \taa uncertainty for \PbPb collisions $\left( ^{+3.4\%}_{-3.9\%}\right)$, respectively.

\begin{figure*}[htb]
  \centering
    \includegraphics[width=0.45\textwidth]{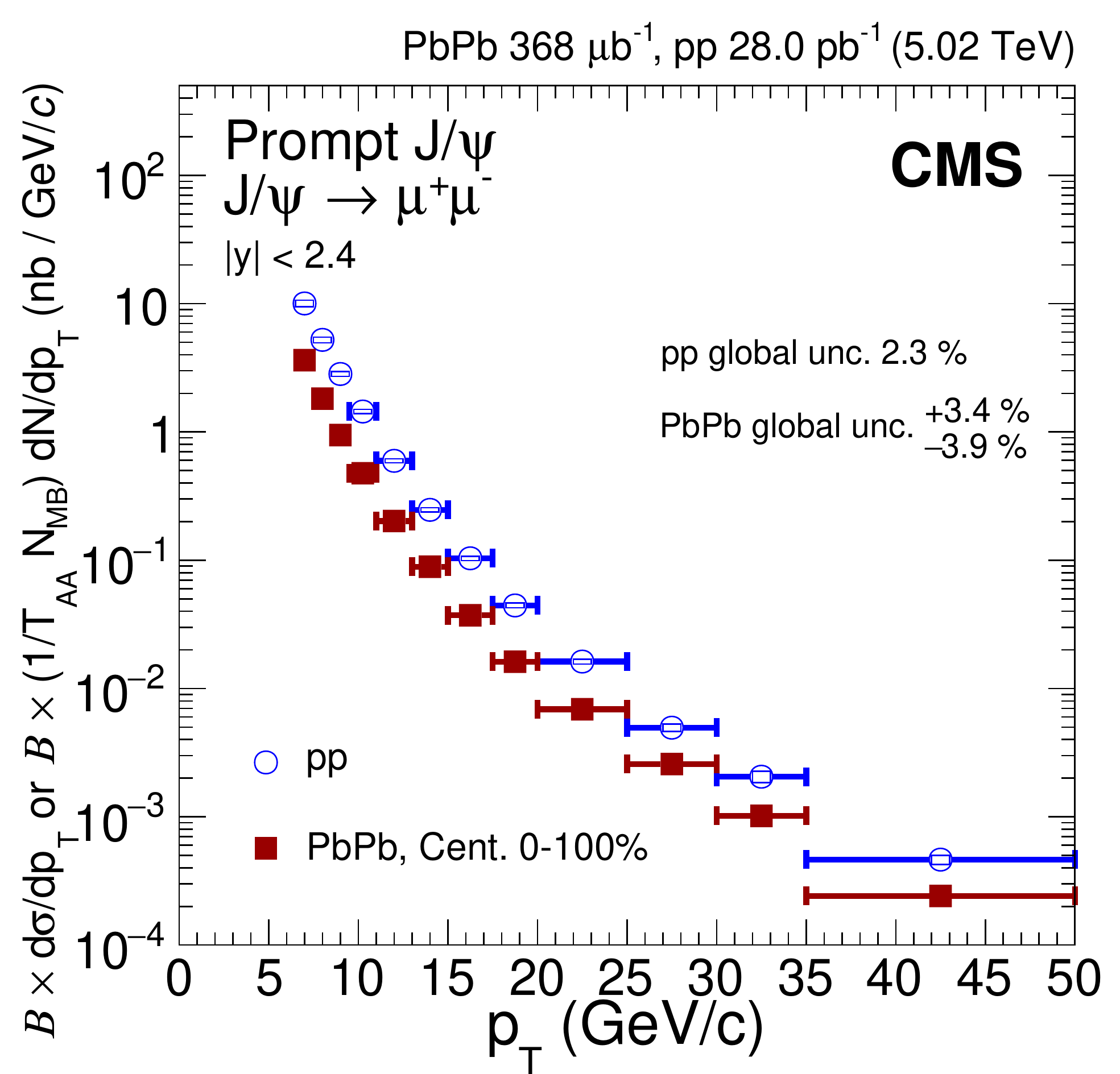}
    \includegraphics[width=0.45\textwidth]{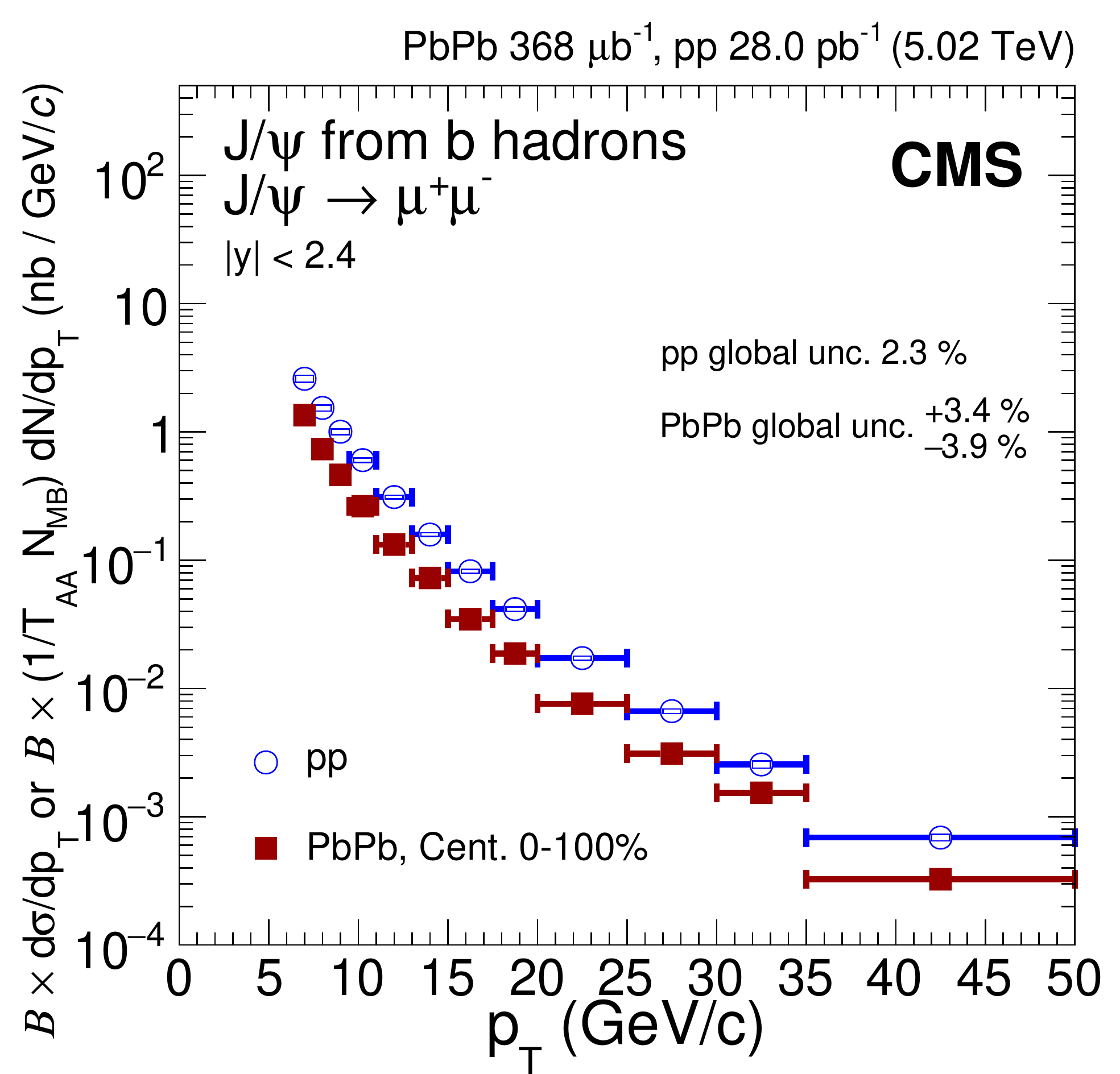}
    \includegraphics[width=0.45\textwidth]{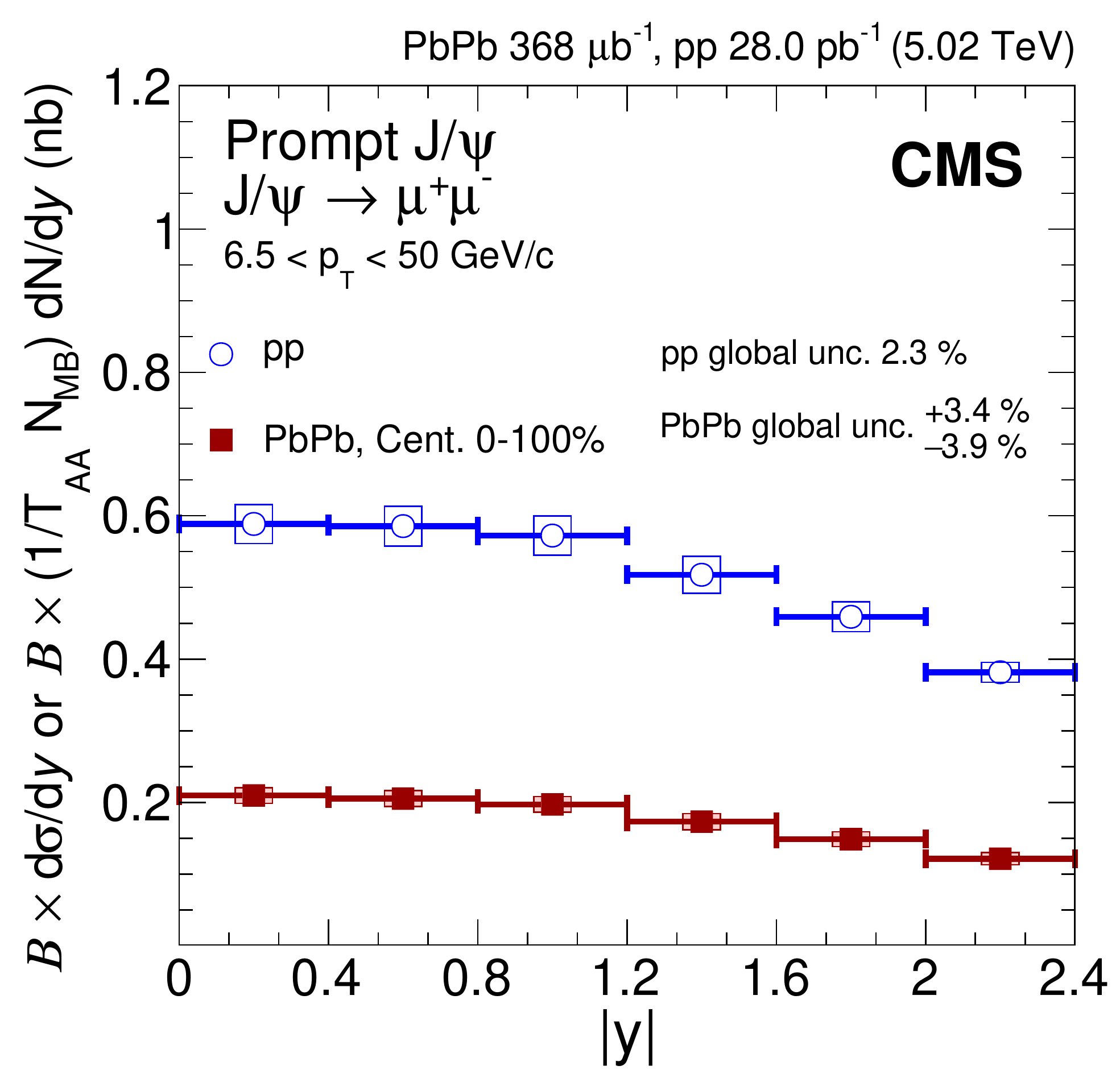}
    \includegraphics[width=0.45\textwidth]{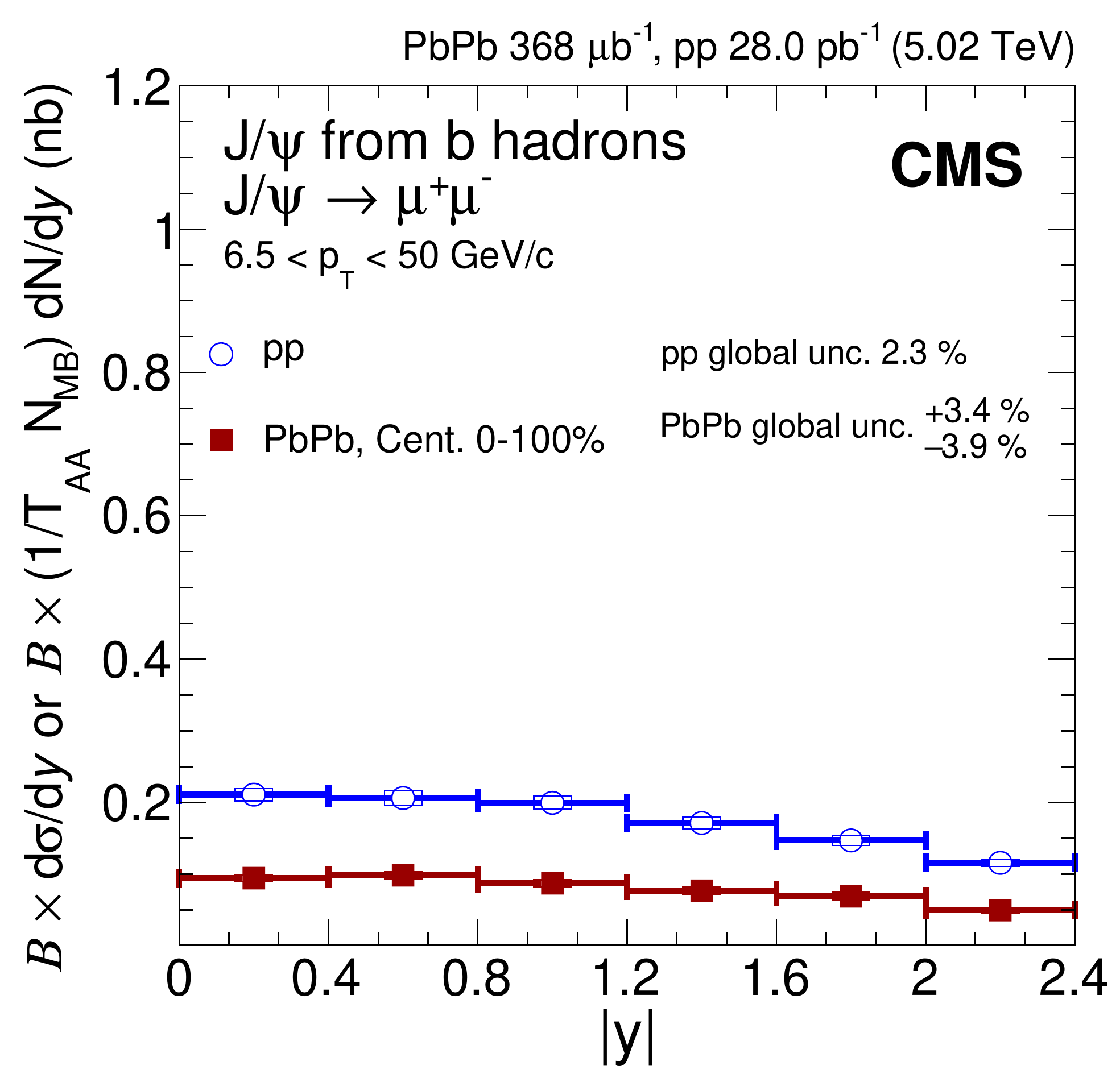}
    \caption{Differential cross section of prompt \JPsi mesons (left) and \JPsi mesons from b hadrons (nonprompt \JPsi) (right)  decaying into two muons as a function of dimuon \pt (upper) and rapidity (lower) in \pp and \PbPb collisions. The \PbPb cross sections are normalised by \taa for direct comparison. The bars (boxes) represent statistical (systematic) point-by-point uncertainties, while global uncertainties are written on the plots.}
    \label{fig:promptJpsi_XS_final_pt}
\end{figure*}

The cross sections for the production of prompt and nonprompt \JPsi mesons that decay into two muons ($\mathcal{B} \sigma$, where $\mathcal{B}$ is the branching ratio of \JPsi to dimuons) are reported as a function of \pt and rapidity in Fig.~\ref{fig:promptJpsi_XS_final_pt}.

\subsection{Prompt \texorpdfstring{\JPsi}{J/psi} meson nuclear modification factor}

In order to compute the nuclear modification factor \raa in a given bin of centrality (cent.), the above-mentioned \PbPb and \pp normalised cross sections are divided in the following way:
\ifthenelse{\boolean{cms@external}}{
\begin{multline*}
   \raa = \frac{N^{\PbPb}_{\JPsi} (\text{cent.}) }{N^{\pp}_{\JPsi}}
   \, \frac{\mathcal{A}^{\pp} \, \epsilon^{\pp}}{\mathcal{A}^{\PbPb} \,  \epsilon^{\PbPb} (\text{cent.}) } \\
   \times \frac{\mathcal{L}^{\pp}}{\nmb \times \langle \taa \rangle \, (\text{cent. fraction})},
\end{multline*}
}{
\begin{equation*}
   \raa = \frac{N^{\PbPb}_{\JPsi} (\text{cent.}) }{N^{\pp}_{\JPsi}}
   \times \frac{\mathcal{A}^{\pp} \times \epsilon^{\pp}}{\mathcal{A}^{\PbPb} \,  \epsilon^{\PbPb} (\text{cent.}) }
   \times \frac{\mathcal{L}^{\pp}}{\nmb \, \langle \taa \rangle \, (\text{cent. fraction})},
\end{equation*}
}
where the centrality fraction is the fraction of the inclusive inelastic cross section probed in the analysis bin. Global uncertainties (indicated as boxes in the plots at $\raa=1$) arise from the full \pp statistical and systematic uncertainties and the \PbPb \nmb uncertainty when binning as a function of the centrality; and from the integrated luminosity of the \pp data, and the \nmb and \taa uncertainties of the \PbPb data, when binning as a function of rapidity or \pt.

\begin{figure*}[htb]
  \centering
    \includegraphics[width=0.45\textwidth]{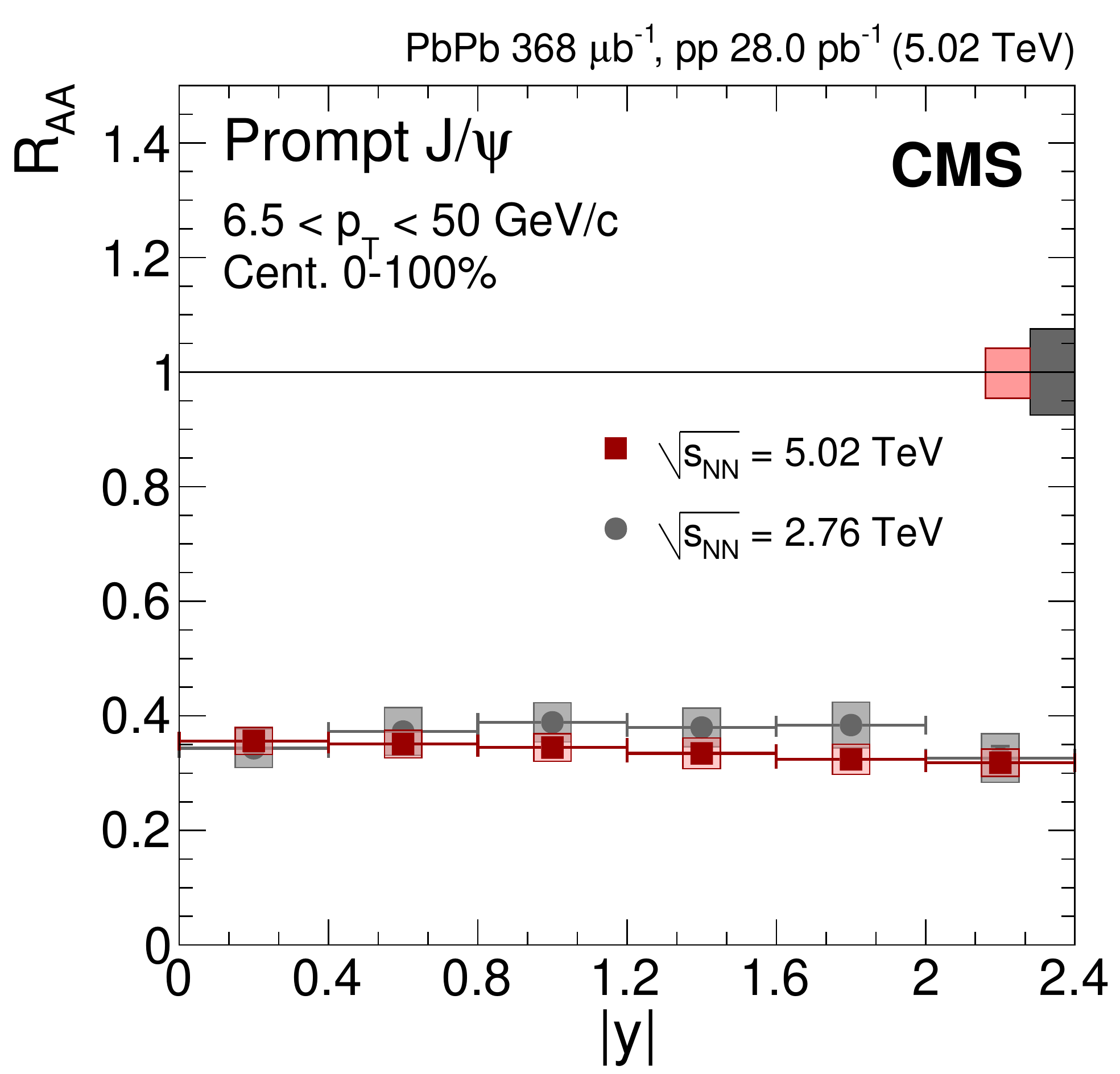}
    \includegraphics[width=0.45\textwidth]{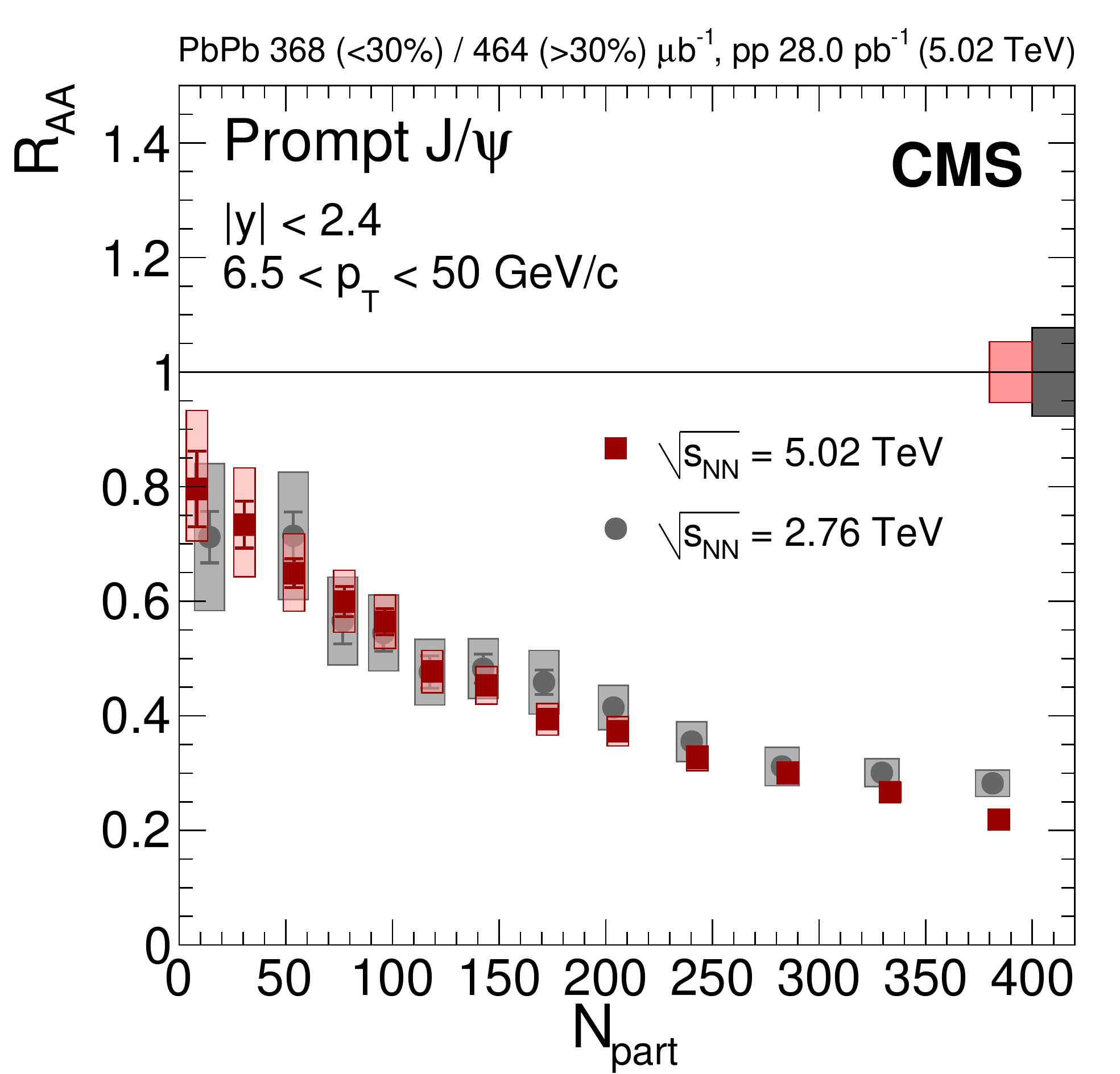}
     \includegraphics[width=0.45\textwidth]{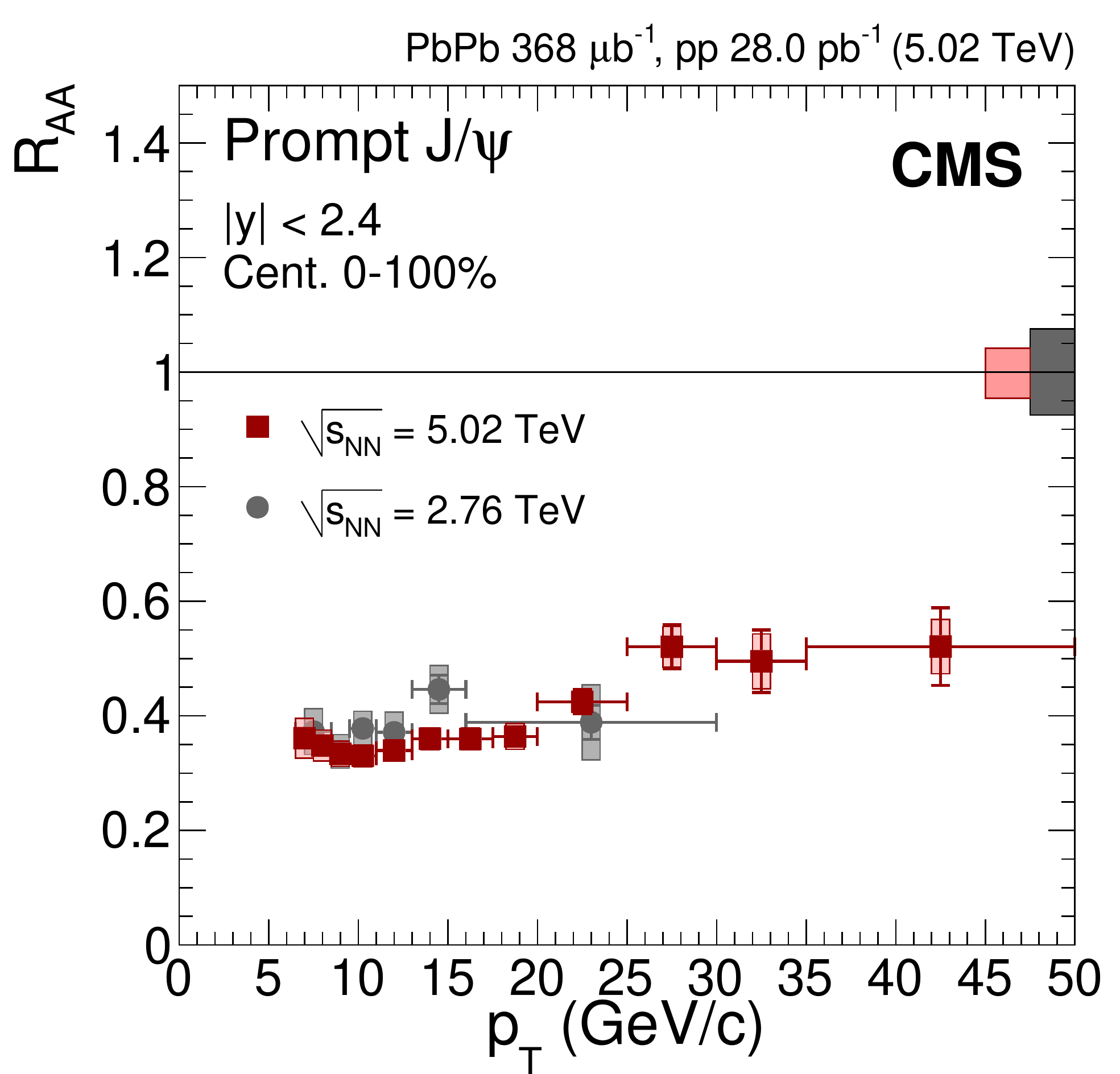}
    \caption{Nuclear modification factor of prompt \JPsi mesons as a function of dimuon rapidity (upper left), \npart (upper right) and dimuon \pt (lower) at $\sqrtsnn = 5.02$\TeV. For the results as a function of \npart the most central bin corresponds to 0--5\%, and the most peripheral one to 70--100\%. Results obtained at 2.76\TeV are overlaid for comparison~\cite{Khachatryan:2016ypw}.
    The bars (boxes) represent statistical (systematic) point-by-point uncertainties. The boxes plotted at $\raa=1$ indicate the size of the global relative uncertainties.
    }
    \label{fig:promptJpsi_RAA_final_allrap}
\end{figure*}

In Fig.~\ref{fig:promptJpsi_RAA_final_allrap}, the \raa of prompt \JPsi mesons
as a function of rapidity, \npart and \pt are shown, integrating in each
case over the other two non-plotted variables.
The results are compared to those obtained at $\sqrtsnn = 2.76$\TeV~\cite{Khachatryan:2016ypw}, and they are found to be in good overall agreement.
No strong rapidity dependence of the suppression is observed.
As a function of centrality, the \raa is suppressed even for the most peripheral bin (70--100\%), with the suppression slowly increasing with \npart. The \raa value for the most central events (0--5\%) is
measured for $6.5<\pt<50$\GeVc and $\abs{y}<2.4$ to be
$0.219 \pm 0.005\stat\pm 0.013\syst$. As a function of \pt the \raa is approximately constant in the range of 5--20\GeVc, but an indication of less suppression at higher \pt is seen for the first time in quarkonia.
Charged hadrons, for which the suppression is usually attributed to parton energy loss~\cite{Arleo:2017ntr,dEnterria:2009xfs}, show a similar increase in \raa at high \pt for \PbPb collisions at $\sqrtsnn = 5.02$\TeV~\cite{Khachatryan:2016odn}.

\begin{figure}[htb]
  \centering
    \includegraphics[width=0.45\textwidth]{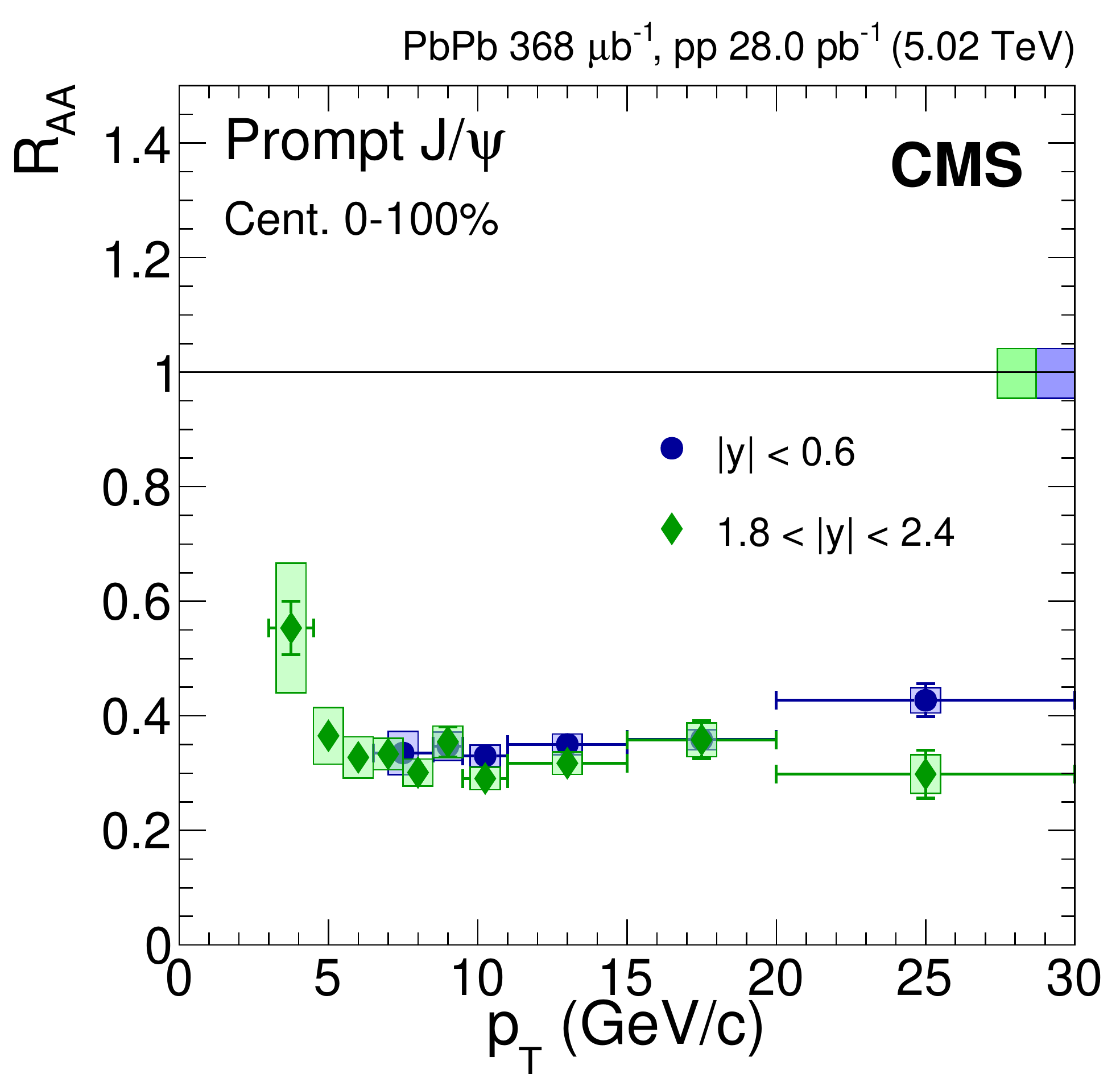}
    \includegraphics[width=0.45\textwidth]{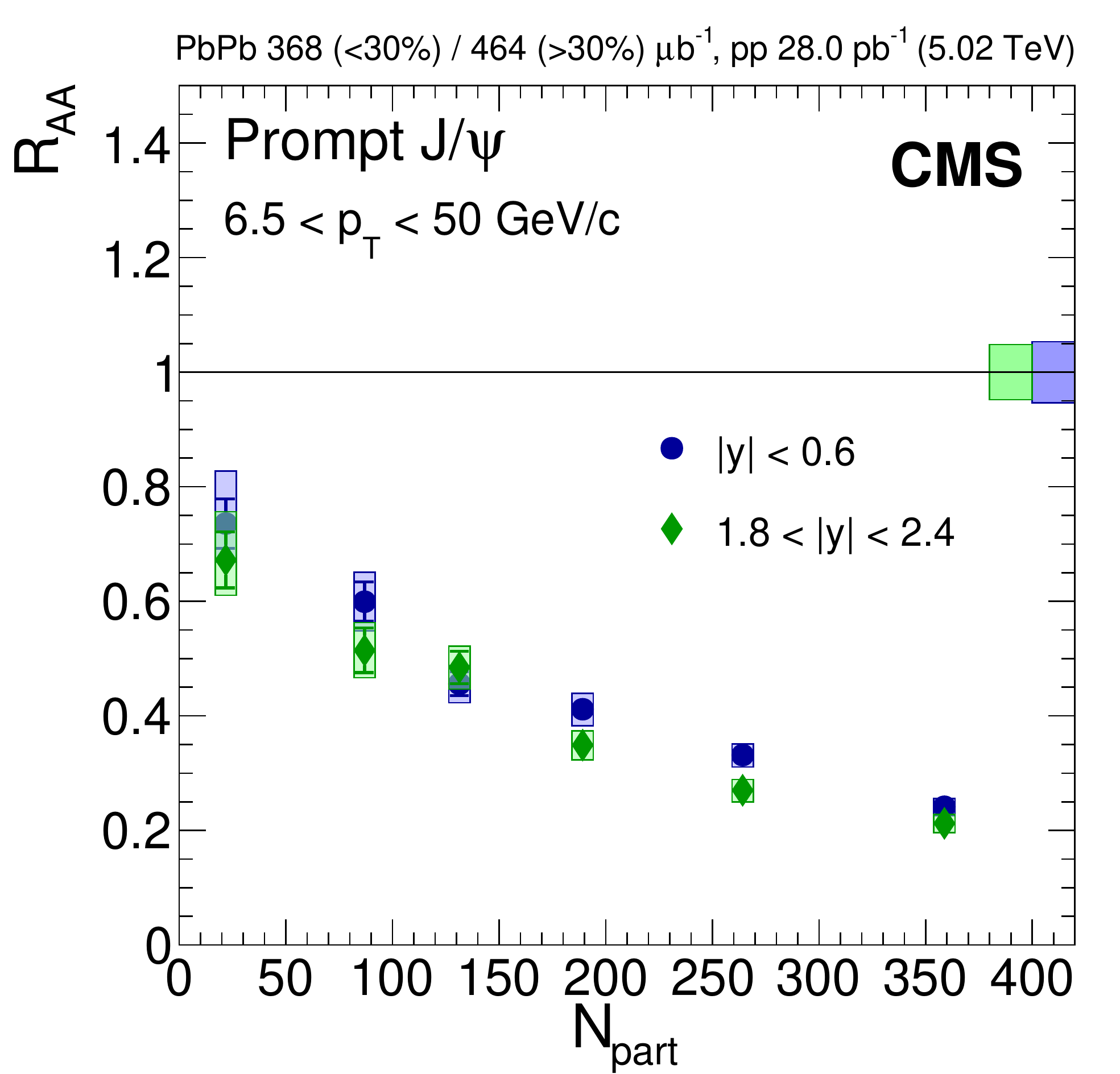}
    \caption{Nuclear modification factor of prompt \JPsi meson as a function of dimuon \pt (\cmsLeft) and  \npart (\cmsRight), in the mid- and most forward rapidity intervals. For the results as a function of \npart the most central bin corresponds to 0--10\%, and the most peripheral one to 50--100\%.
     The bars (boxes) represent statistical (systematic) point-by-point uncertainties. The boxes plotted at $\raa=1$ indicate the size of the global relative uncertainties.
           }
    \label{fig:promptJpsi_RAA_final_rapbins}
\end{figure}

\begin{figure}[htb]
  \centering
    \includegraphics[width=0.45\textwidth]{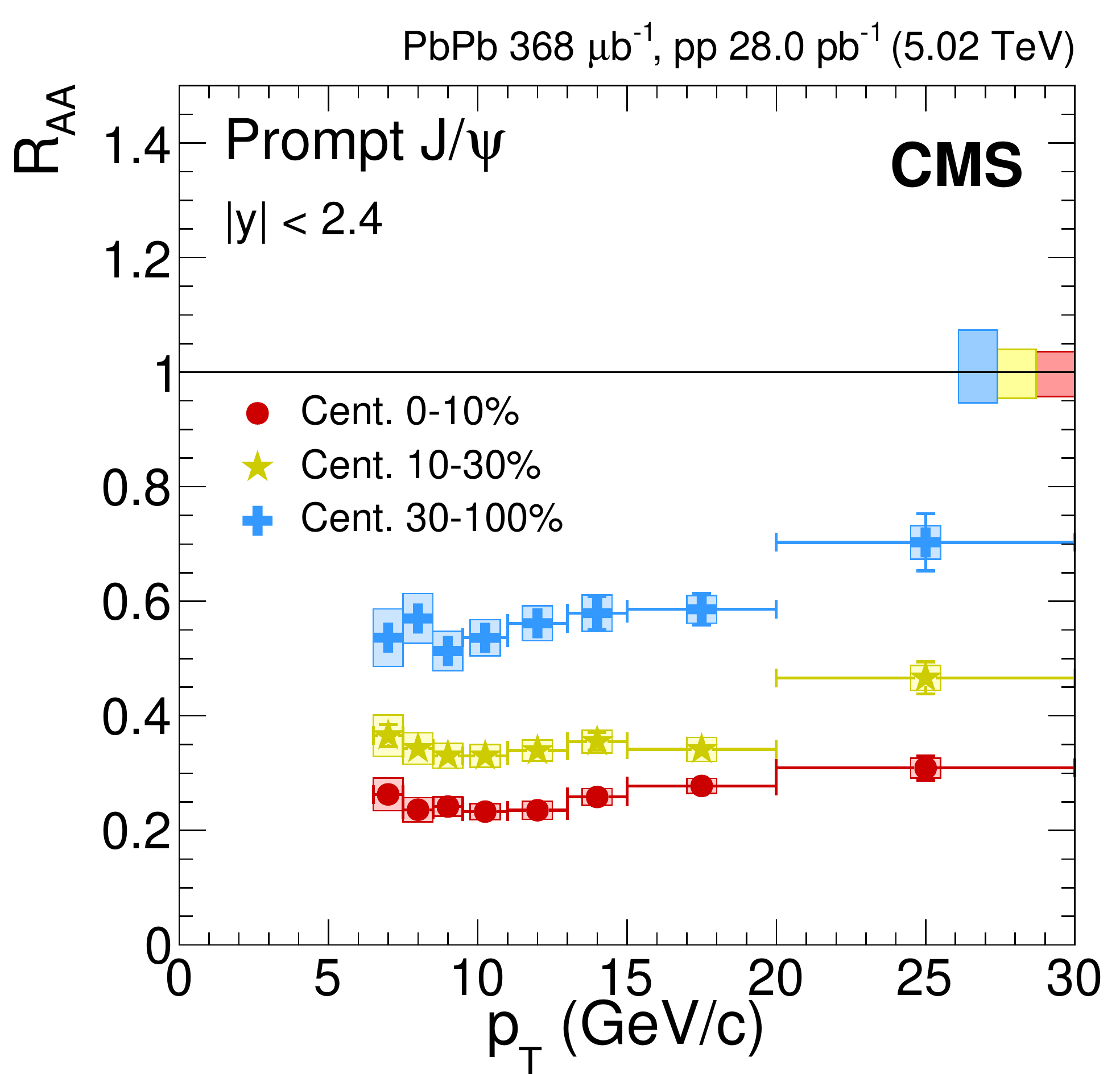}
        \includegraphics[width=0.45\textwidth]{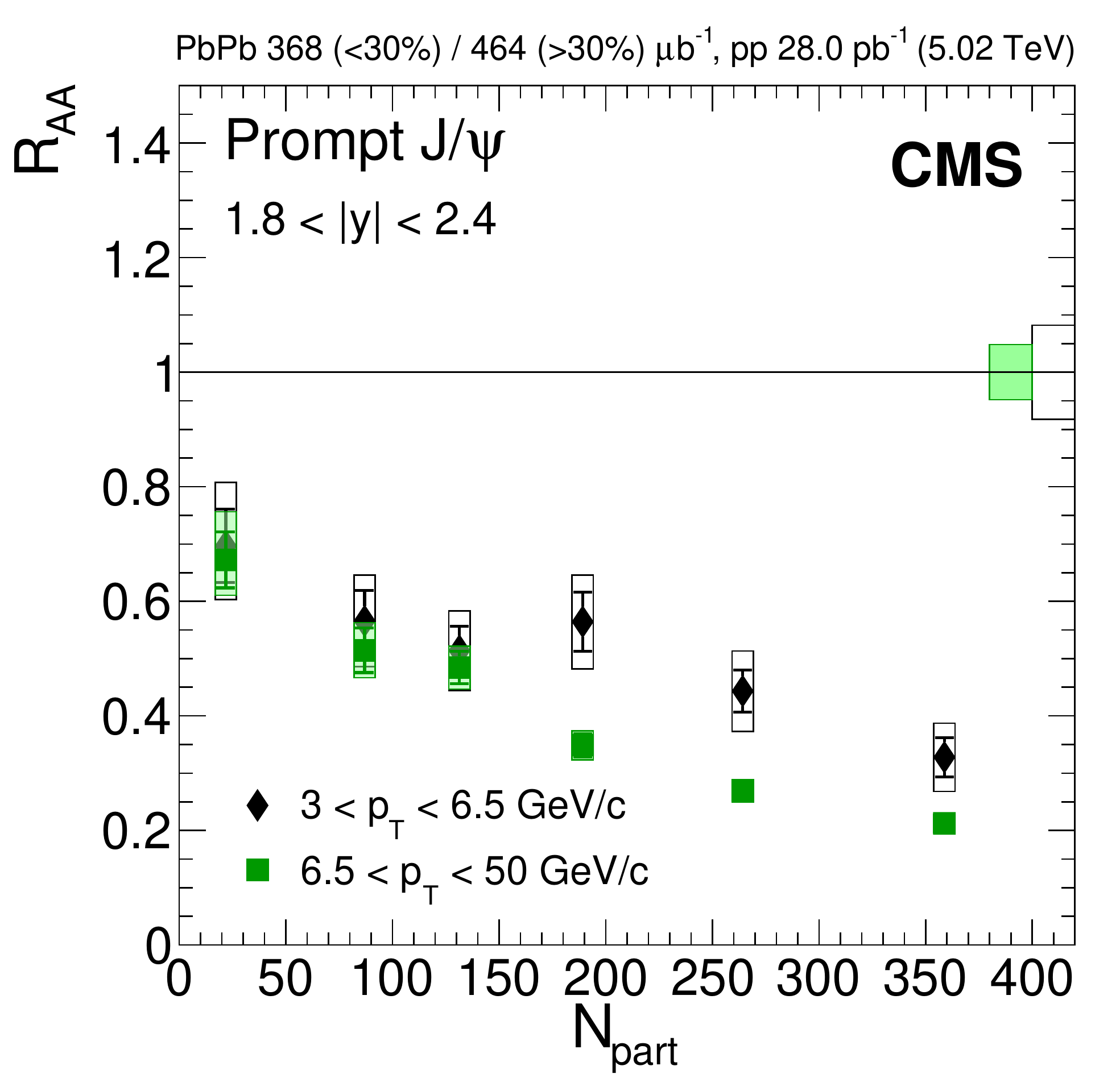}
    \caption{Nuclear modification factor of prompt \JPsi mesons. \cmsLLeft: as a function of dimuon \pt in three centrality bins. \cmsRRight: as a function of \npart at moderate and high \pt, in the forward $1.8<\abs{y}<2.4$ range. For the results as a function of \npart the most central bin corresponds to 0--10\%, and the most peripheral one to 50--100\%. The bars (boxes) represent statistical (systematic) point-by-point uncertainties. The boxes plotted at $\raa=1$ indicate the size of the global relative uncertainties.
           }
    \label{fig:promptJpsi_RAA_final_centbins}
\end{figure}

Double-differential studies are also performed. Figure~\ref{fig:promptJpsi_RAA_final_rapbins} shows the \pt (\cmsLeft) and centrality (\cmsRight) dependence of prompt \JPsi \raa measured in the mid- and most forward rapidity intervals. A similar suppression pattern is observed for both rapidities. Figure~\ref{fig:promptJpsi_RAA_final_centbins} (\cmsLeft) shows the dependence of \raa as a function of \pt, for three centrality intervals. Although the mean level of suppression strongly depends on the sampled centrality range, the general trend of the \pt dependence appears similar in all three centrality ranges, including the increase of \raa at high \pt. Finally, Fig.~\ref{fig:promptJpsi_RAA_final_centbins} (\cmsRight) considers the rapidity interval $1.8<\abs{y}<2.4$, where the acceptance goes down at lower \pt. The suppression is found to be similar in peripheral events at moderate ($3<\pt<6.5$\GeVc) and high ($6.5<\pt<50$\GeVc) transverse momentum ranges, but it is weaker for lower \pt in the most central region. This is also reflected in the first bin of the most forward measurement in Fig.~\ref{fig:promptJpsi_RAA_final_rapbins} (\cmsLeft). A similarly reduced suppression at low \pt is observed by the ALICE Collaboration, which is attributed to a regeneration contribution~\cite{Abelev:2012rv,Adam:2016rdg}.

\subsection{Prompt \texorpdfstring{\psiP}{psi(2S)} meson nuclear modification factor}

Having measured the prompt \JPsi \raa, one can derive that of the \psiP meson by multiplying it by the double ratio $(N_{\psiP}/N_{\JPsi})_{\PbPb}/ (N_{\psiP}/N_{\JPsi})_{\pp}$ of the relative modification of the prompt \psiP and \JPsi meson yields from \pp to \PbPb collisions published in Ref.~\cite{HIN16004}. Since the \psiP yield suffers from lower statistics, the current \JPsi analysis is repeated using the wider bins of Ref.~\cite{HIN16004}. The centrality binning used is 0--10--20--30--40--50--100\% for the results in $\abs{y}<1.6$, and 0--20--40--100\% for the results in $1.6<\abs{y}<2.4$. Since the statistical uncertainty in the \psiP largely dominates, the \JPsi uncertainties are propagated by considering them to be uncorrelated to the double ratio uncertainties.

\begin{figure*}[htb]
  \centering
    \includegraphics[width=0.45\textwidth]{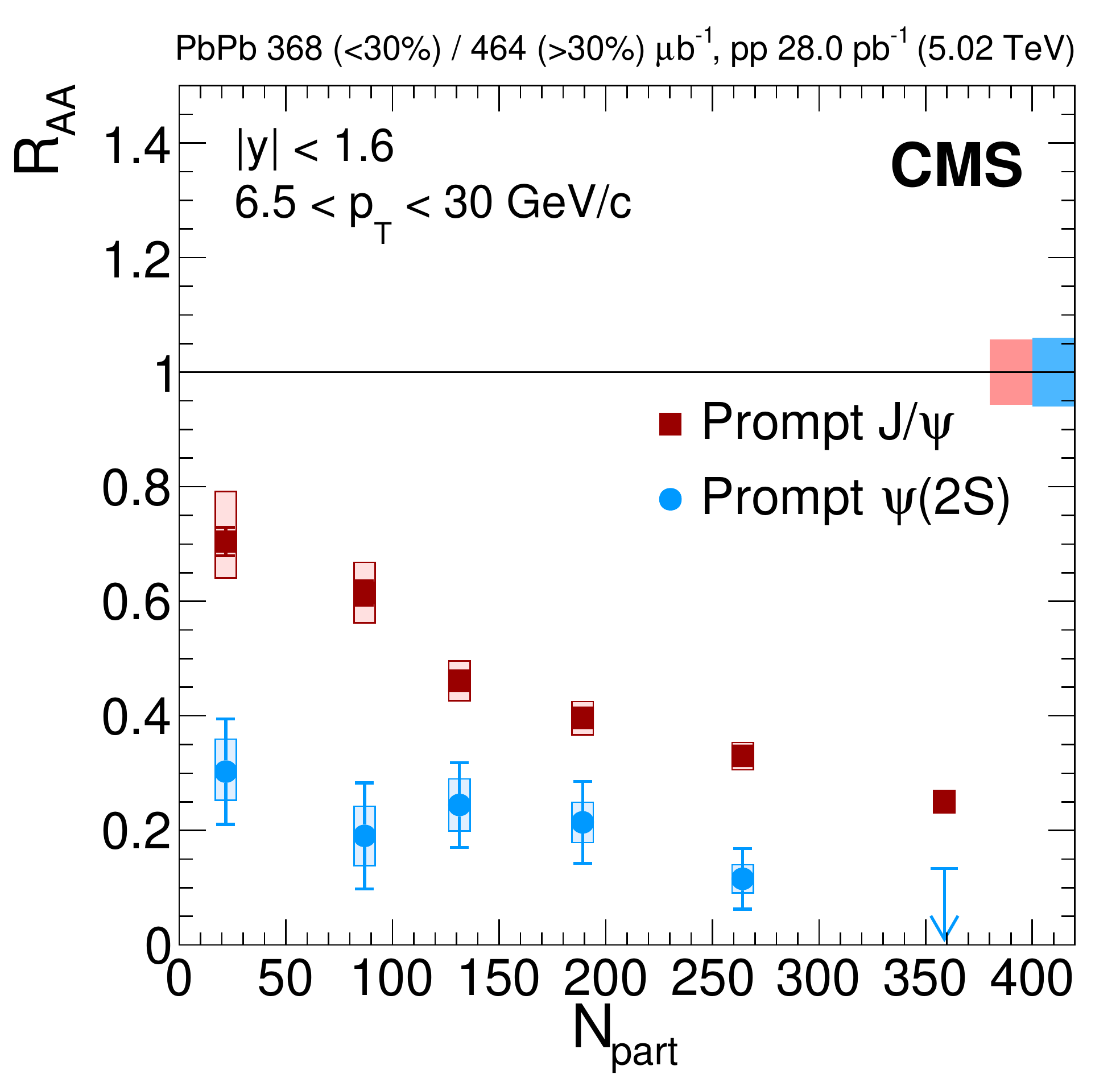}
    \includegraphics[width=0.45\textwidth]{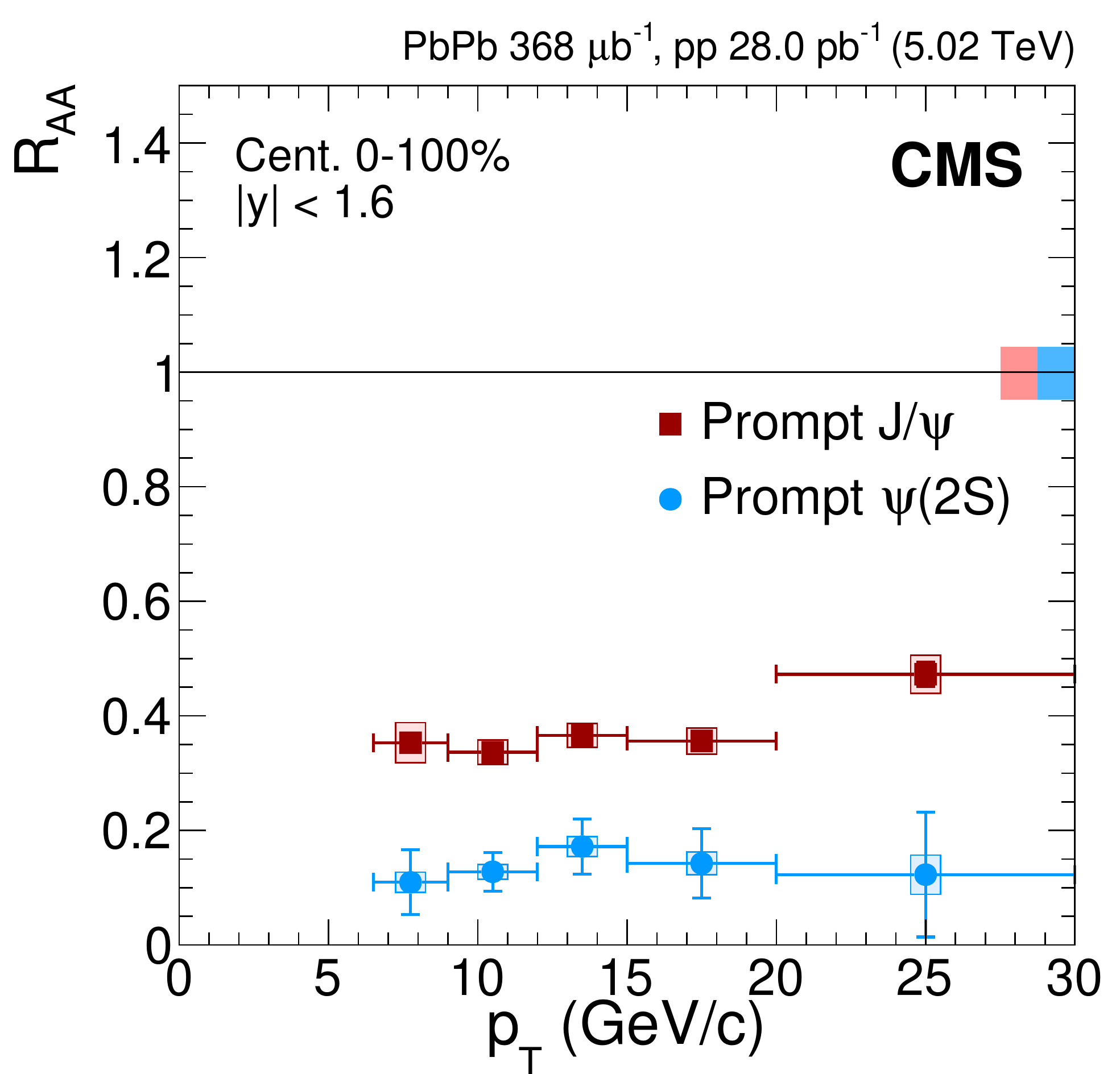}
    \includegraphics[width=0.45\textwidth]{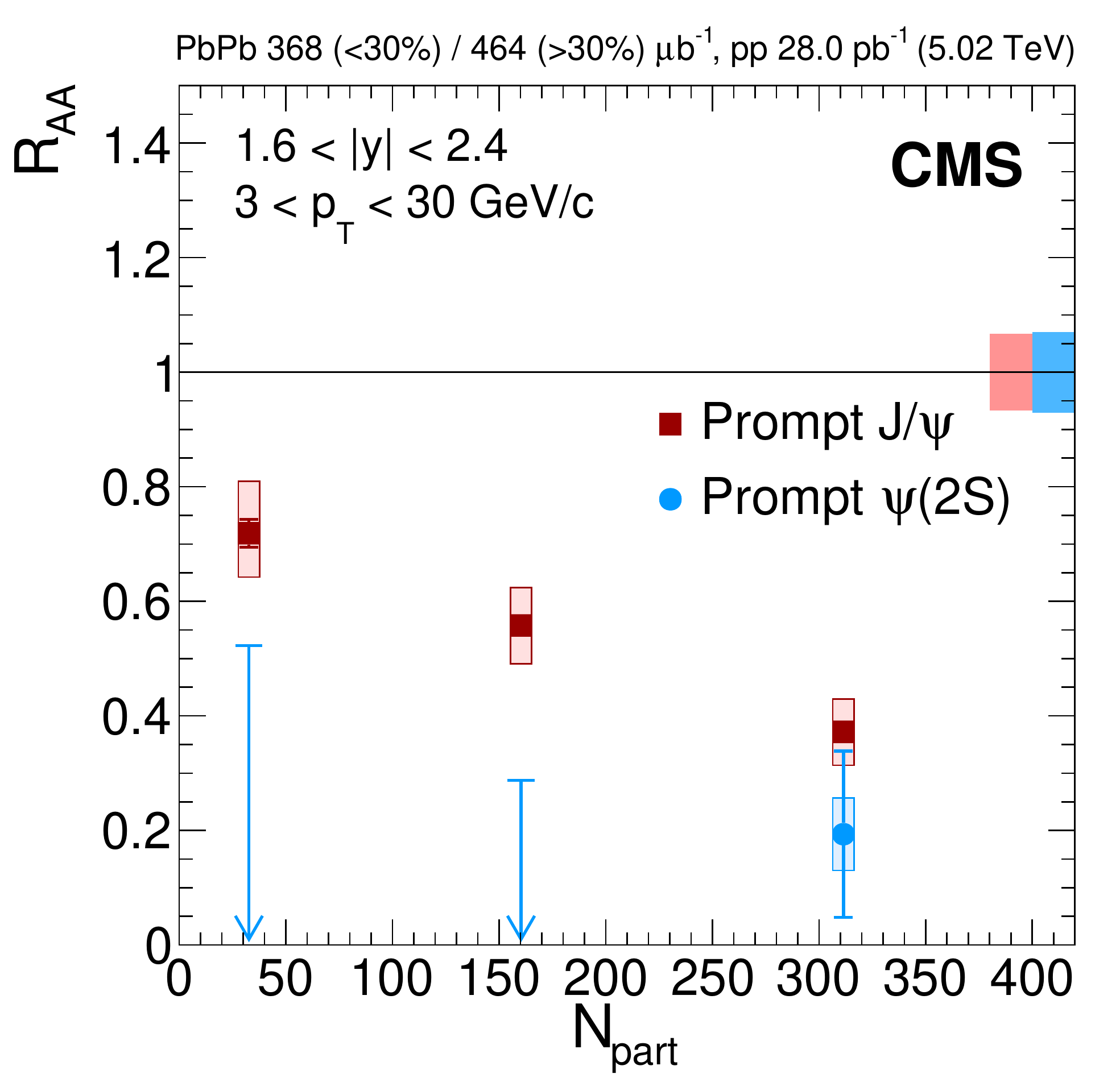}
    \includegraphics[width=0.45\textwidth]{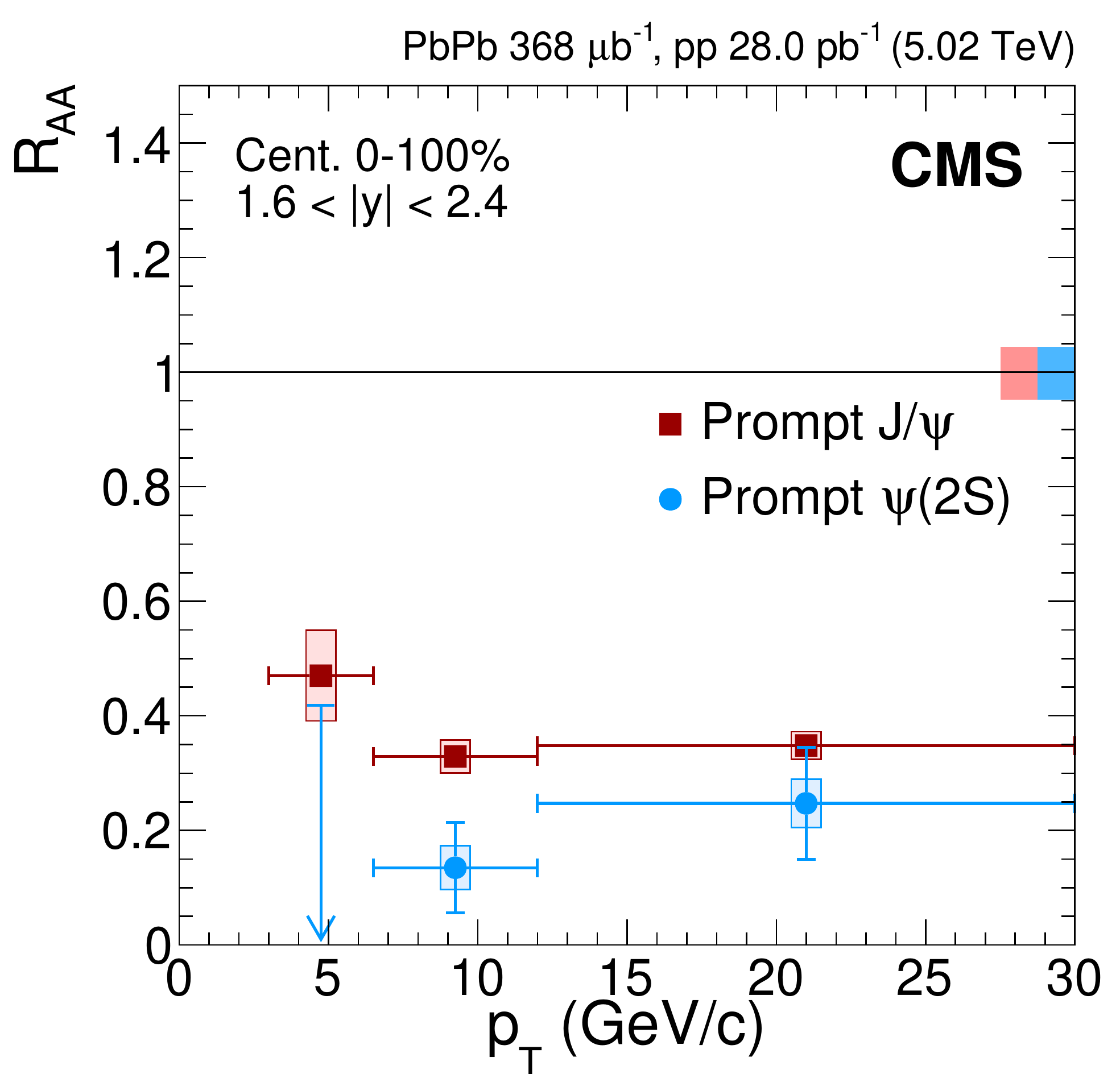}
    \caption{Nuclear modification factor of prompt \JPsi and \psiP mesons as a function of \npart (left) and dimuon \pt (right), at central (upper, starting at $\pt=6.5$\GeVc) and forward (lower, starting at $\pt=3.0$\GeVc) rapidity. The vertical arrows represent 95\% confidence intervals in the bins where the double ratio measurement is consistent with 0 (see text). For the results as a function of \npart the most central bin corresponds to 0--10\% (0--20\%), and the most peripheral one to 50--100\% (40--100\%), for $\abs{y}<1.6$ ($1.6<\abs{y}<2.4$). The bars (boxes) represent statistical (systematic) point-by-point uncertainties. The boxes plotted at $\raa = 1$ indicate the size of the global relative uncertainties.}
    \label{fig:promptPsiP_RAA_final_ptcentbins}
\end{figure*}

The results are presented in Fig.~\ref{fig:promptPsiP_RAA_final_ptcentbins} as a function of dimuon \pt and \npart, in two rapidity ranges of different \pt reach.
In the bins where the double ratio is consistent with 0, 95\% CL intervals on the prompt \psiP \raa are derived using the Feldman--Cousins procedure~\cite{Feldman:1997qc}. The procedure to obtain the CL intervals is the same as in the double ratio measurement, incorporating the \JPsi \raa statistical and systematic uncertainties as a nuisance parameter.
It can be observed that the \psiP meson production is more suppressed than that of \JPsi mesons, in the entire measured range. The \psiP meson \raa shows no clear dependence of the suppression with \pt, and hints of an increasing suppression with collision centrality. These results show that the \psiP mesons are more strongly affected by the medium created in \PbPb collisions than the \JPsi mesons.

\subsection{Nonprompt \texorpdfstring{\JPsi}{J/psi} meson nuclear modification factor}

\begin{figure*}[htb]
  \centering
    \includegraphics[width=0.45\textwidth]{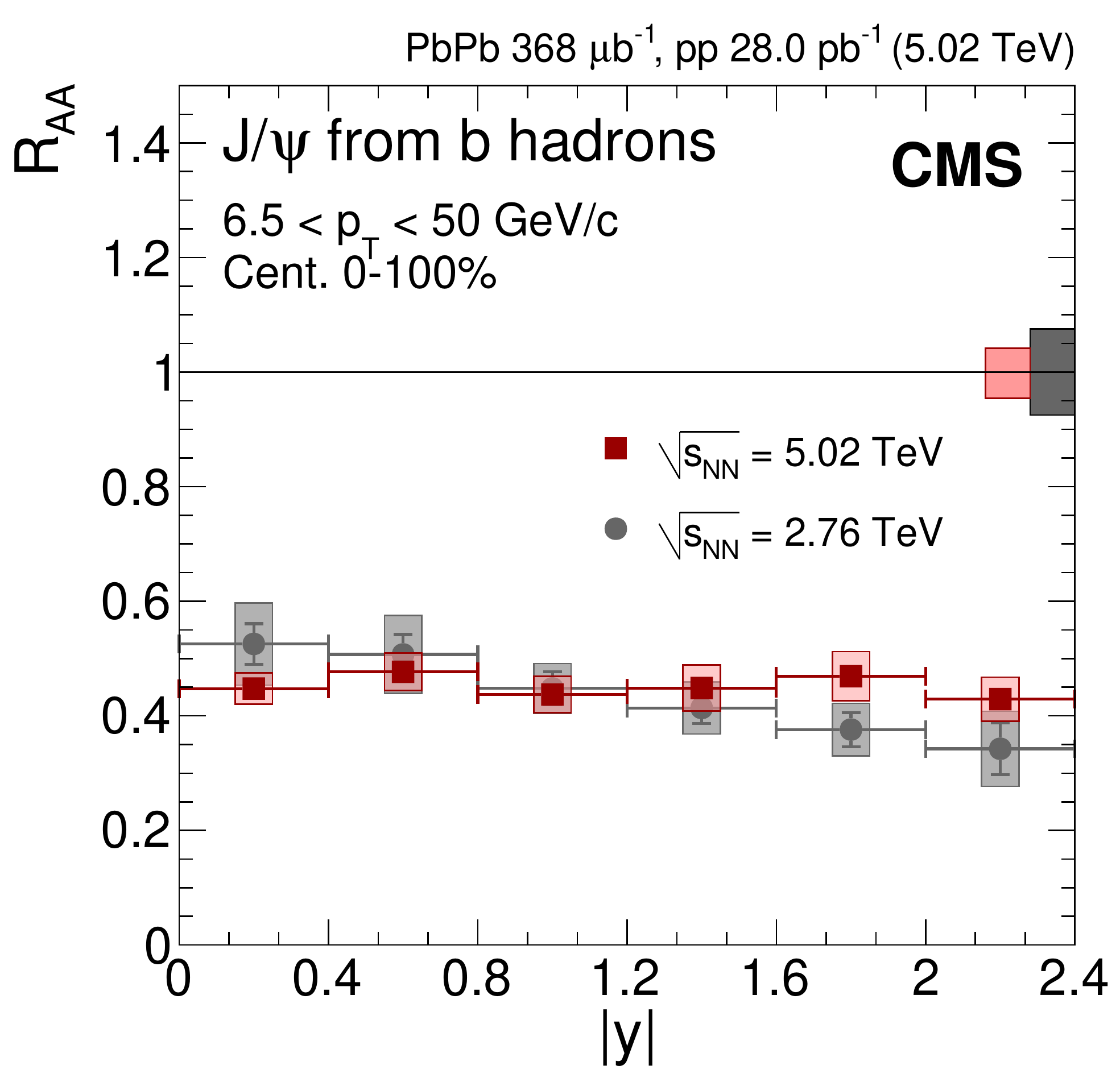}
    \includegraphics[width=0.45\textwidth]{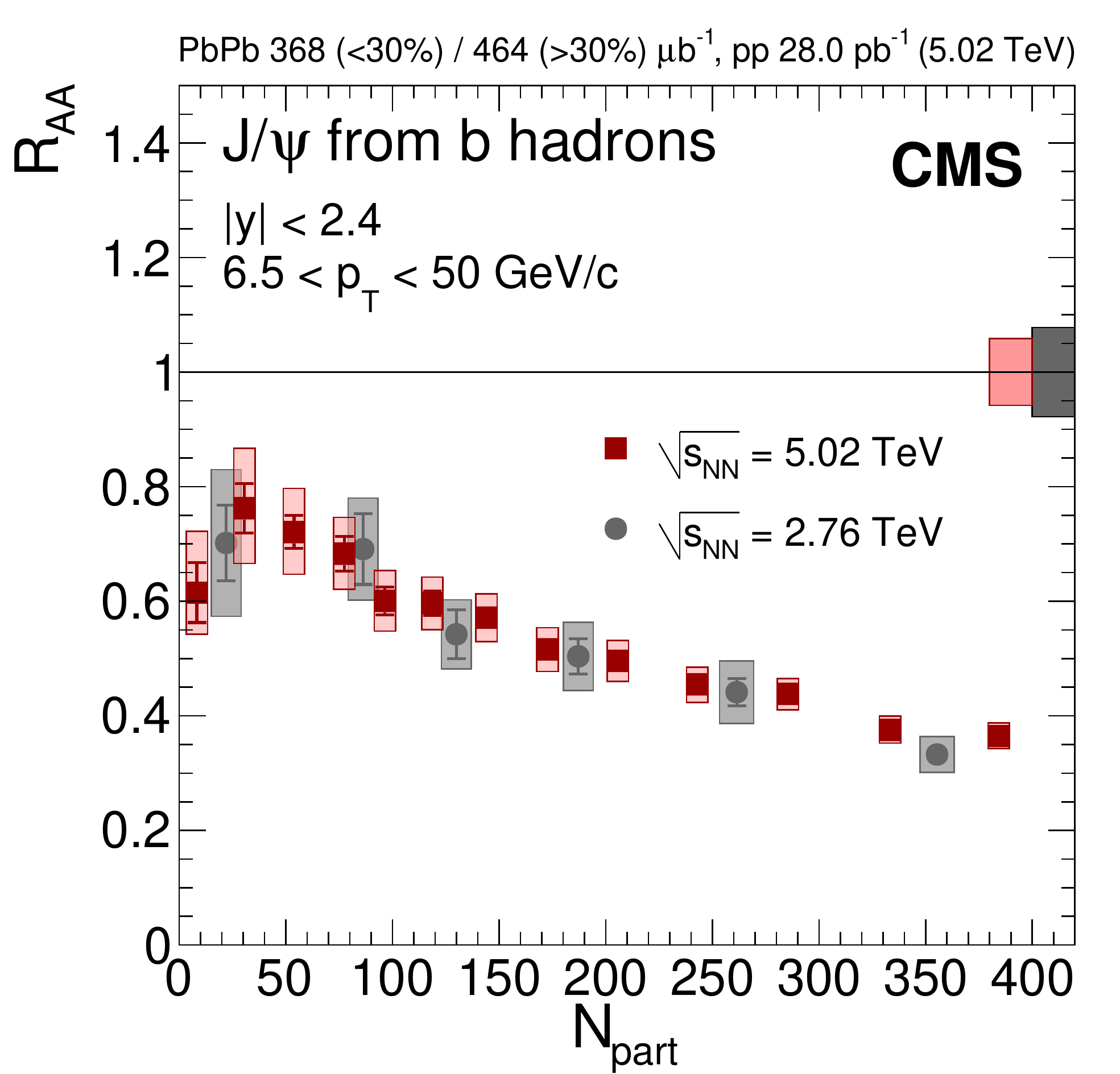}
     \includegraphics[width=0.45\textwidth]{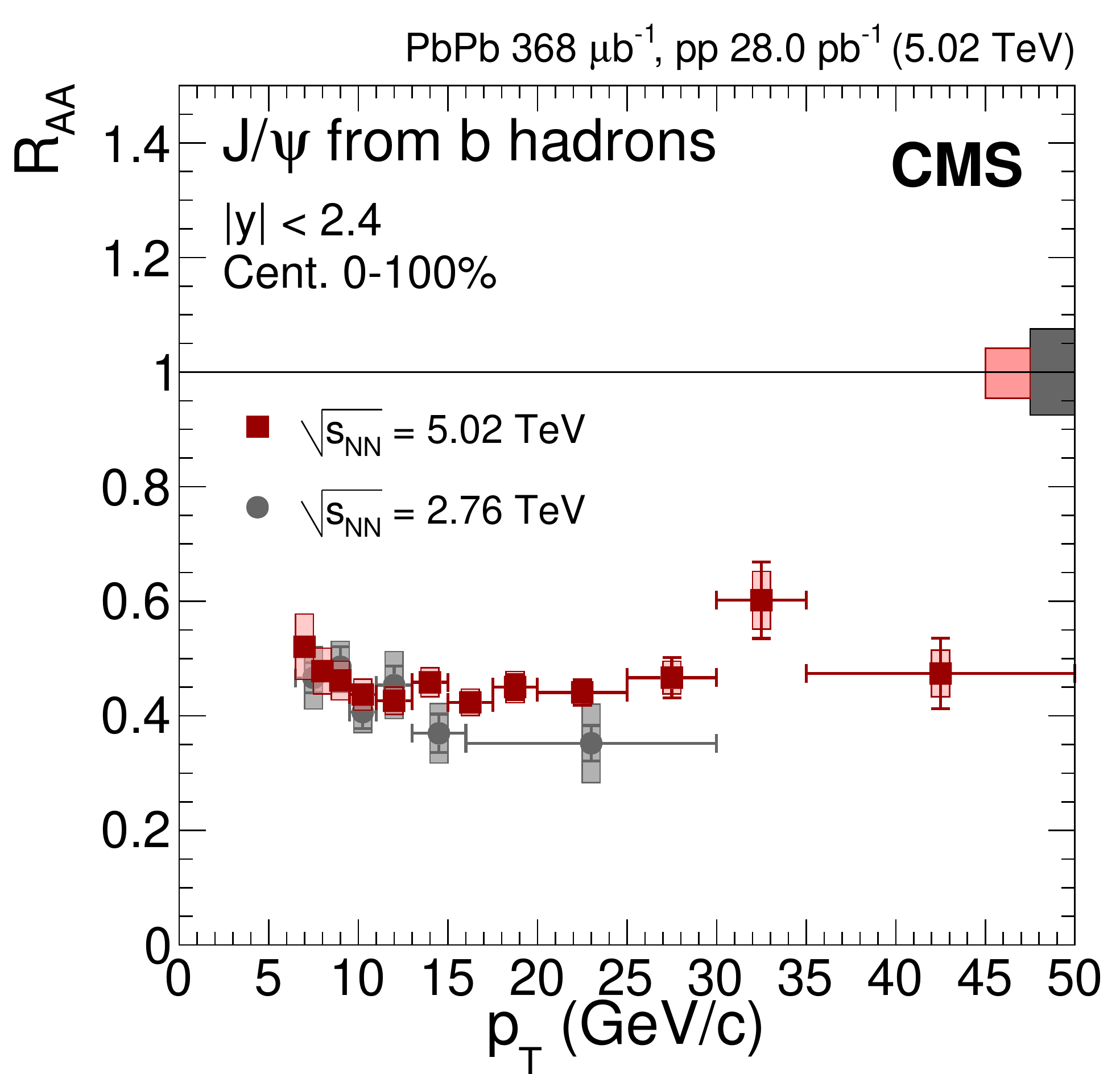}
    \caption{Nuclear modification factor of \JPsi mesons from b hadrons (nonprompt \JPsi) as a function of dimuon rapidity (upper left), \npart (upper right) and dimuon \pt (lower)~at $\sqrtsnn = 5.02$\TeV. For the results as a function of \npart the most central bin corresponds to 0--5\%, and the most peripheral one to 70--100\%. Results obtained at 2.76\TeV are overlaid for comparison~\cite{Khachatryan:2016ypw}.
The bars (boxes) represent statistical (systematic) point-by-point uncertainties. The boxes plotted at $\raa=1$ indicate the size of the global relative uncertainties.
    }
    \label{fig:nonpromptJpsi_RAA_final_allrap}
\end{figure*}

The procedure applied to derive the prompt \JPsi meson \raa is applied to the nonprompt component. In Fig.~\ref{fig:nonpromptJpsi_RAA_final_allrap}, the \raa of nonprompt \JPsi
as a function of rapidity, centrality and \pt are shown, integrating in each
case over the other two non-plotted variables. The results are compared to those obtained at $\sqrtsnn = 2.76$\TeV~\cite{Khachatryan:2016ypw}. A good overall agreement is found, although no rapidity dependence is observed in the present analysis, while the suppression was slowly increasing towards forward rapidities in the lower-energy measurement.
A steady increase of the suppression is observed with increasing centrality of
the collision. The \raa for the most central events (0--5\%) measured for
$6.5<\pt<50$\GeVc and $\abs{y}<2.4$ is $0.365\pm 0.009\stat\pm0.022\syst$.

\begin{figure}[htb]
  \centering
    \includegraphics[width=0.45\textwidth]{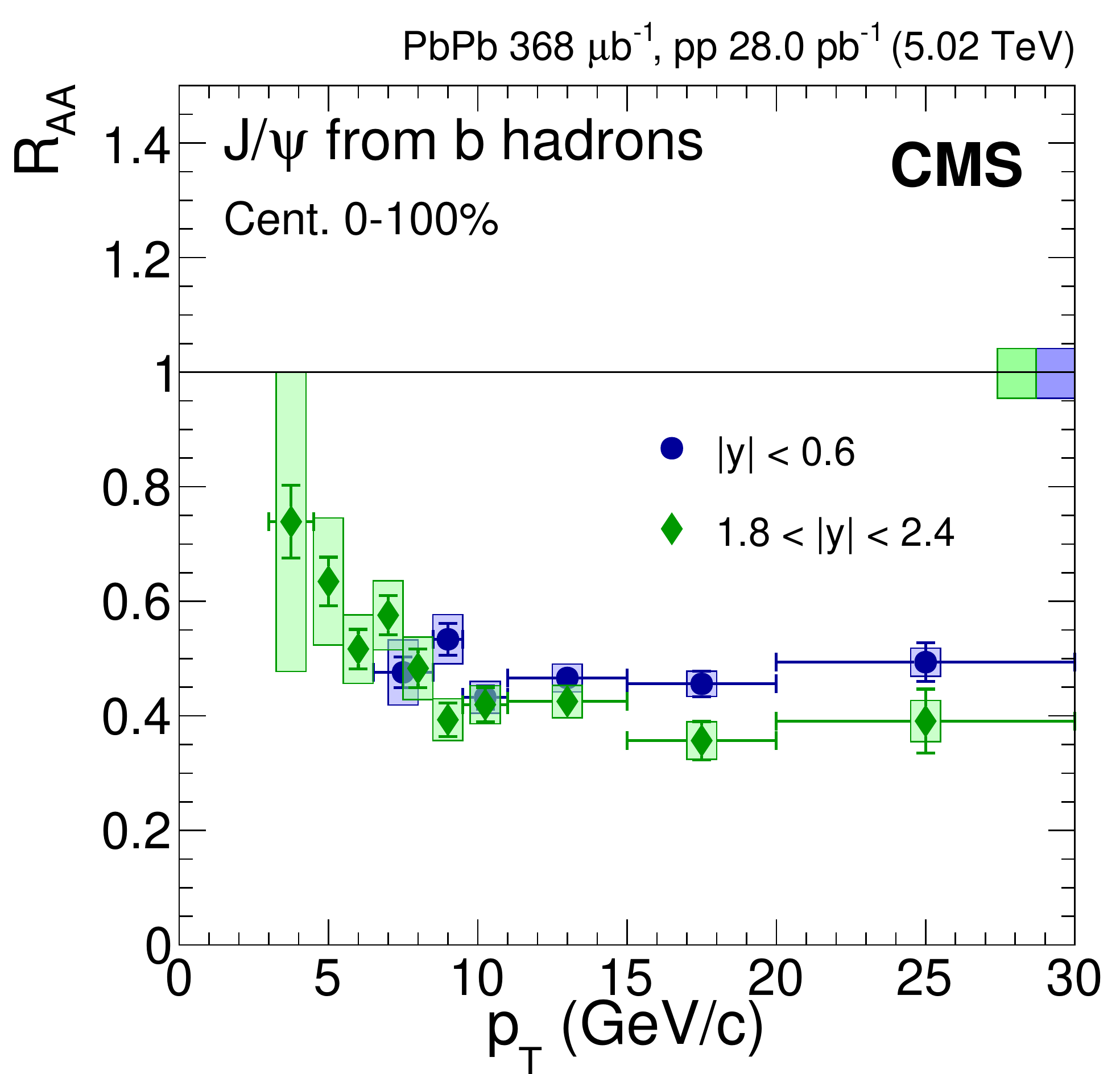}
    \includegraphics[width=0.45\textwidth]{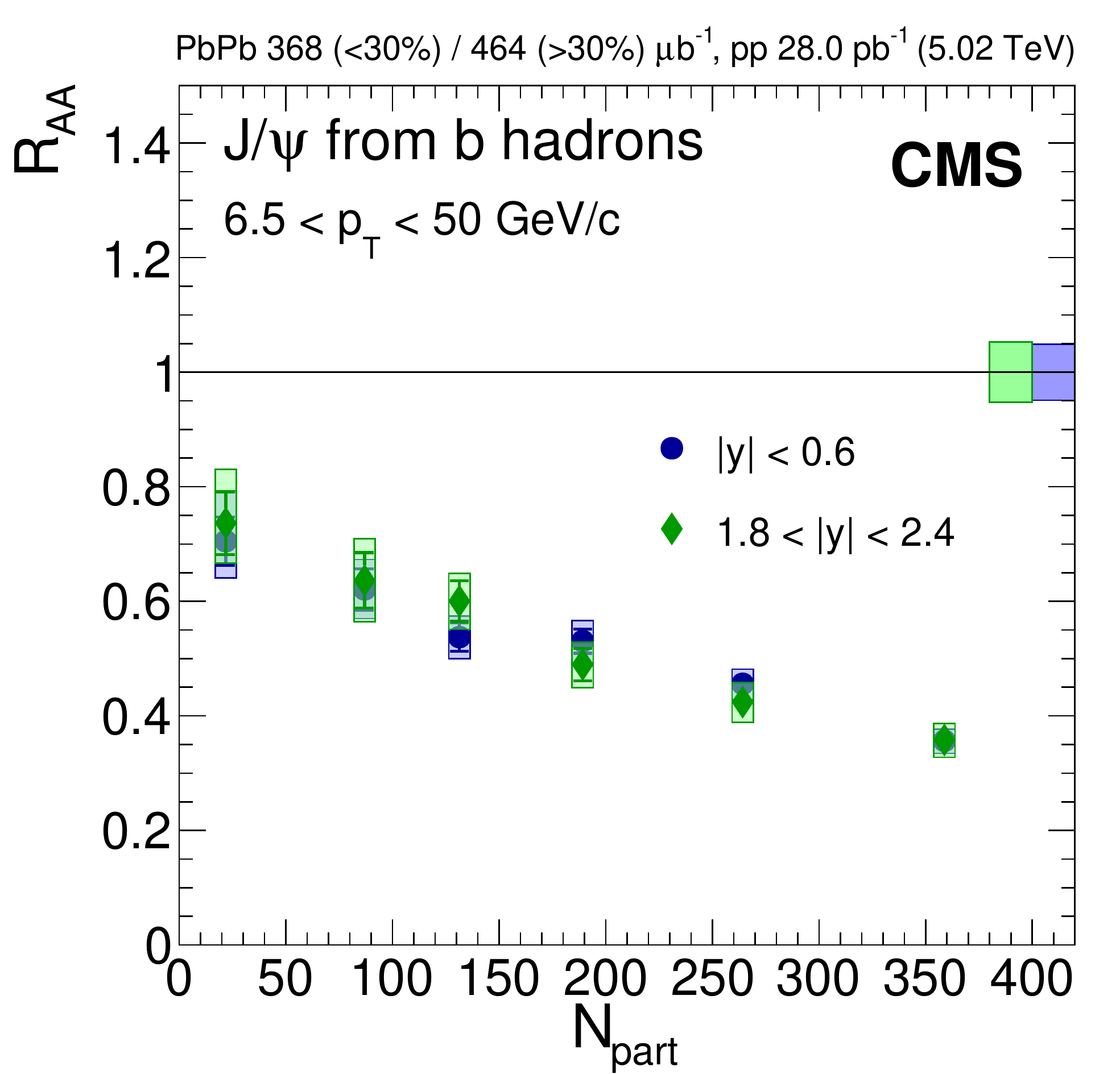}
    \caption{Nuclear modification factor of \JPsi mesons from b hadrons (nonprompt \JPsi) as a function of dimuon \pt (\cmsLeft) and \npart (\cmsRight) and in the mid- and most forward rapidity intervals. For the results as a function of \npart the most central bin corresponds to 0--10\%, and the most peripheral one to 50--100\%.
     The bars (boxes) represent statistical (systematic) point-by-point
           uncertainties. The boxes plotted at $\raa=1$ indicate the size of the global relative uncertainties.
           }
    \label{fig:nonpromptJpsi_RAA_final_rapbins}
\end{figure}

\begin{figure}[htb]
  \centering
    \includegraphics[width=0.45\textwidth]{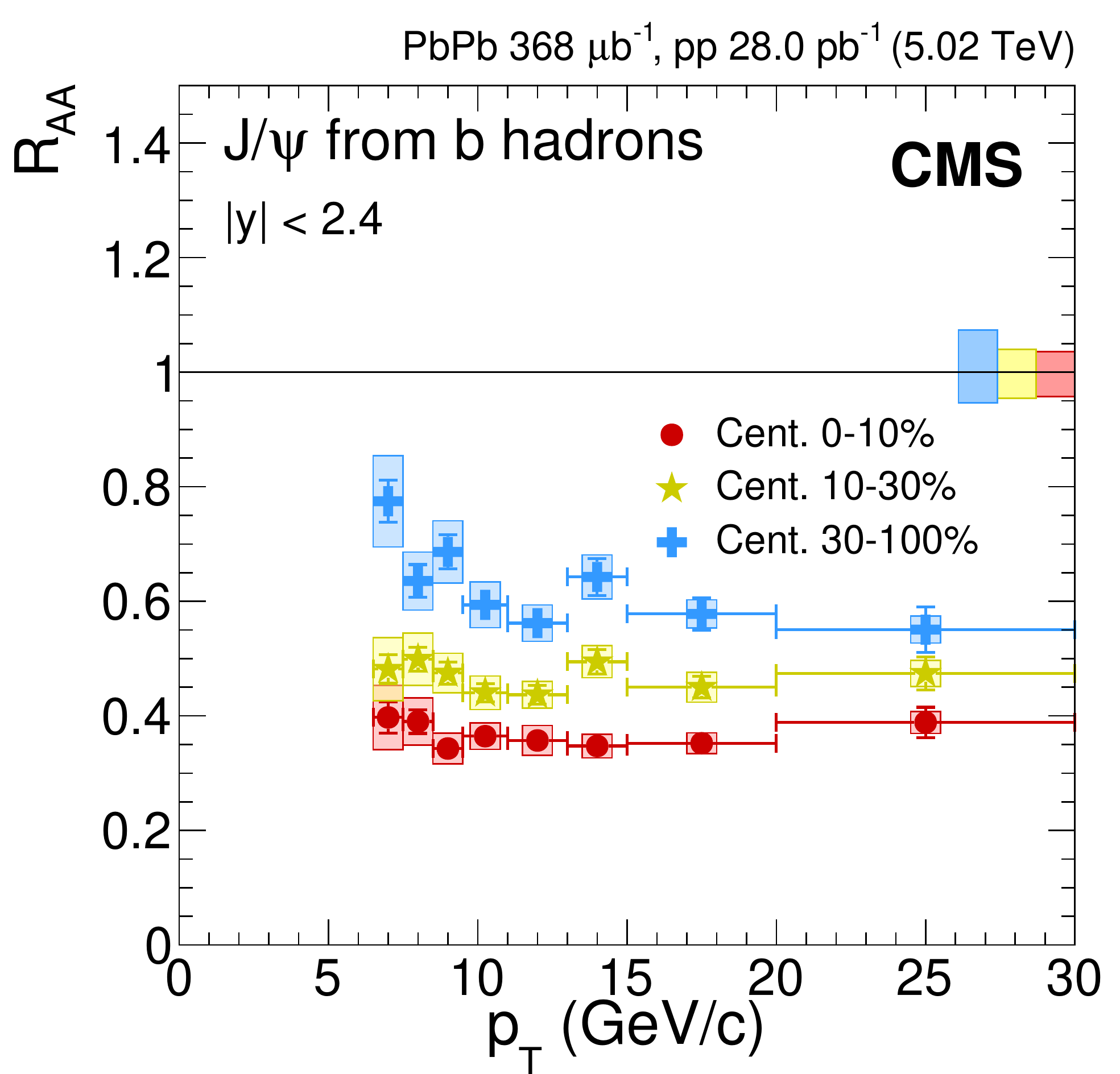}
    \includegraphics[width=0.45\textwidth]{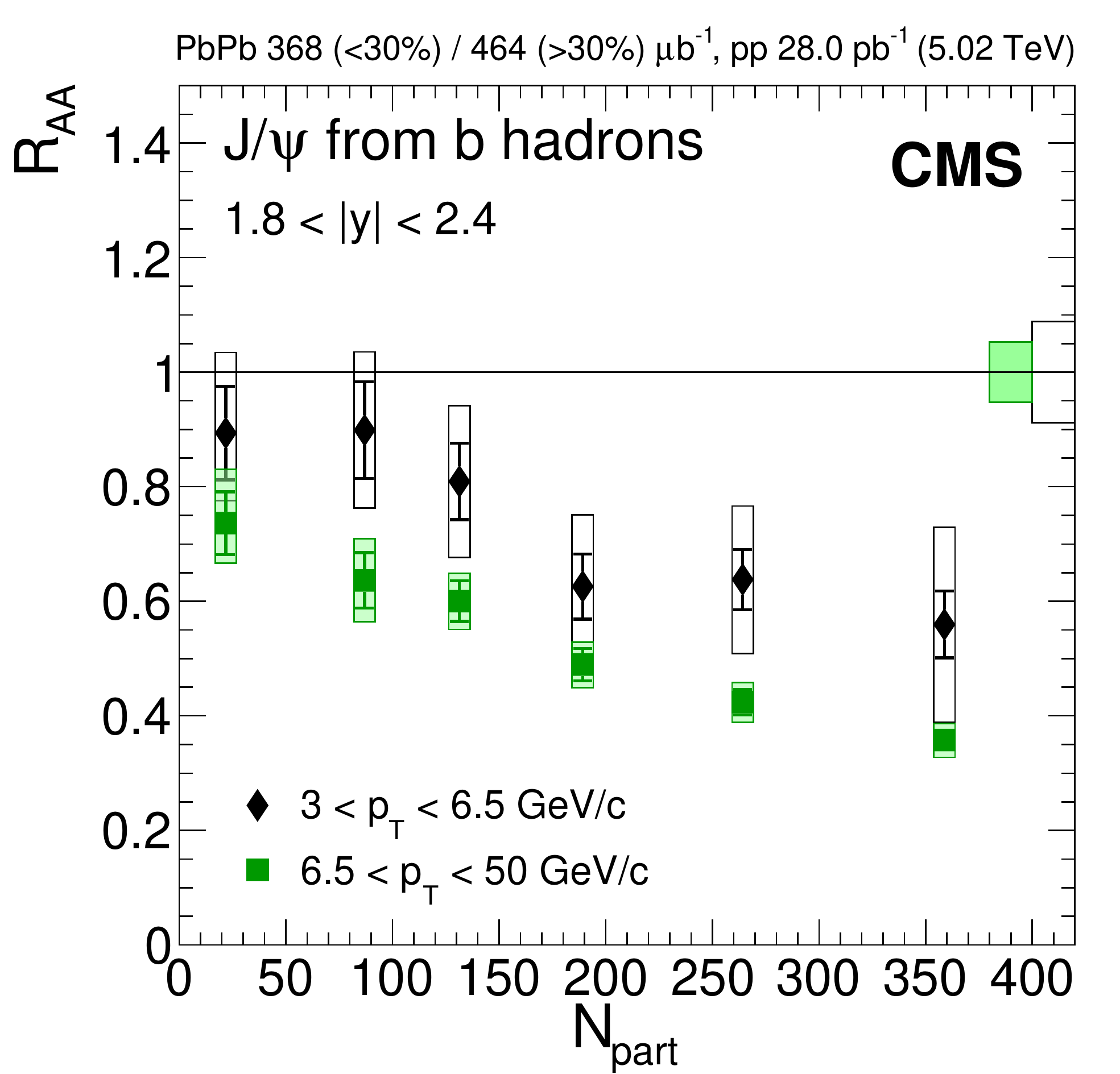}
    \caption{Nuclear modification factor of \JPsi mesons from b hadrons (nonprompt \JPsi). \cmsLLeft: as a function of dimuon \pt in three centrality bins. \cmsRRight: as a function of \npart at moderate and high \pt, in the forward $1.8<\abs{y}<2.4$ range. For the results as a function of \npart the most central bin corresponds to 0--10\%, and the most peripheral one to 50--100\%. The bars (boxes) represent statistical (systematic) point-by-point uncertainties. The boxes plotted at $\raa=1$ indicate the size of the global relative uncertainties.
           }
    \label{fig:nonpromptJpsi_RAA_final_centbins}
\end{figure}

As for the prompt production case, double-differential studies are also performed.
Figure~\ref{fig:nonpromptJpsi_RAA_final_rapbins} shows the \pt (\cmsLeft) and centrality (\cmsRight) dependence of nonprompt \JPsi meson \raa measured in the mid- and most forward rapidity intervals. No strong rapidity dependence is observed, and a hint of a smaller suppression at low \pt is seen in the $1.8<\abs{y}<2.4$ range. Figure~\ref{fig:nonpromptJpsi_RAA_final_centbins} (\cmsLeft) shows the dependence of \raa as a function of \pt, for three centrality ranges. While the nonprompt \JPsi meson \raa does not seem to depend on rapidity, the data indicates a larger \pt dependence in peripheral events.
Finally, Fig.~\ref{fig:nonpromptJpsi_RAA_final_centbins} (\cmsRight) shows, for $1.8<\abs{y}<2.4$, \raa as a function of \npart, for two \pt intervals. Hints of a stronger suppression are seen for $\pt>6.5\GeVc$ at all centralities.

 \section{Conclusions}

Prompt and nonprompt \JPsi meson production has been studied in \pp and \PbPb collisions at $\sqrtsnn = 5.02\TeV$, as a function of rapidity, transverse momentum (\pt), and collision centrality, in different kinematic and centrality ranges.
Three observables were measured: nonprompt \JPsi fractions, prompt and nonprompt \JPsi cross sections for each collision system, and nuclear modification factors \raa.
The \raa results show a strong centrality dependence, with an increasing suppression for increasing centrality. For both prompt and nonprompt \JPsi mesons no significant dependence on rapidity is observed. An indication of less suppression in the lowest \pt range at forward rapidity is seen for both \JPsi components. Double-differential measurements show the same trend, and also suggest a stronger \pt dependence in peripheral events. An indication of less suppression of the prompt \JPsi meson at high \pt is seen with respect to that observed at intermediate \pt. The measurements are consistent with previous results at $\sqrtsnn = 2.76\TeV$.

Combined with previous results for the double ratio $(N_{\psiP}/N_{\JPsi})_{\PbPb}/ (N_{\psiP}/N_{\JPsi})_{\pp}$, the current \raa values for \JPsi mesons are used to derive the prompt \psiP meson \raa in \PbPb collisions at $\sqrtsnn = 5.02\TeV$, as a function of \pt and collision centrality, in two different rapidity ranges.
The results show that the \psiP is more suppressed than the \JPsi meson for all the kinematical ranges studied. No \pt dependence is observed within the current uncertainties. Hints of an increase in suppression with increasing collision centrality are also observed.

\begin{acknowledgments}
We congratulate our colleagues in the CERN accelerator departments for the excellent performance of the LHC and thank the technical and administrative staffs at CERN and at other CMS institutes for their contributions to the success of the CMS effort. In addition, we gratefully acknowledge the computing centres and personnel of the Worldwide LHC Computing Grid for delivering so effectively the computing infrastructure essential to our analyses. Finally, we acknowledge the enduring support for the construction and operation of the LHC and the CMS detector provided by the following funding agencies: BMWFW and FWF (Austria); FNRS and FWO (Belgium); CNPq, CAPES, FAPERJ, and FAPESP (Brazil); MES (Bulgaria); CERN; CAS, MoST, and NSFC (China); COLCIENCIAS (Colombia); MSES and CSF (Croatia); RPF (Cyprus); SENESCYT (Ecuador); MoER, ERC IUT, and ERDF (Estonia); Academy of Finland, MEC, and HIP (Finland); CEA and CNRS/IN2P3 (France); BMBF, DFG, and HGF (Germany); GSRT (Greece); OTKA and NIH (Hungary); DAE and DST (India); IPM (Iran); SFI (Ireland); INFN (Italy); MSIP and NRF (Republic of Korea); LAS (Lithuania); MOE and UM (Malaysia); BUAP, CINVESTAV, CONACYT, LNS, SEP, and UASLP-FAI (Mexico); MBIE (New Zealand); PAEC (Pakistan); MSHE and NSC (Poland); FCT (Portugal); JINR (Dubna); MON, RosAtom, RAS, RFBR and RAEP (Russia); MESTD (Serbia); SEIDI, CPAN, PCTI and FEDER (Spain); Swiss Funding Agencies (Switzerland); MST (Taipei); ThEPCenter, IPST, STAR, and NSTDA (Thailand); TUBITAK and TAEK (Turkey); NASU and SFFR (Ukraine); STFC (United Kingdom); DOE and NSF (USA).

\hyphenation{Rachada-pisek} Individuals have received support from the Marie-Curie programme and the European Research Council and Horizon 2020 Grant, contract No. 675440 (European Union); the Leventis Foundation; the A. P. Sloan Foundation; the Alexander von Humboldt Foundation; the Belgian Federal Science Policy Office; the Fonds pour la Formation \`a la Recherche dans l'Industrie et dans l'Agriculture (FRIA-Belgium); the Agentschap voor Innovatie door Wetenschap en Technologie (IWT-Belgium); the Ministry of Education, Youth and Sports (MEYS) of the Czech Republic; the Council of Science and Industrial Research, India; the HOMING PLUS programme of the Foundation for Polish Science, cofinanced from European Union, Regional Development Fund, the Mobility Plus programme of the Ministry of Science and Higher Education, the National Science Center (Poland), contracts Harmonia 2014/14/M/ST2/00428, Opus 2014/13/B/ST2/02543, 2014/15/B/ST2/03998, and 2015/19/B/ST2/02861, Sonata-bis 2012/07/E/ST2/01406; the National Priorities Research Program by Qatar National Research Fund; the Programa Severo Ochoa del Principado de Asturias; the Thalis and Aristeia programmes cofinanced by EU-ESF and the Greek NSRF; the Rachadapisek Sompot Fund for Postdoctoral Fellowship, Chulalongkorn University and the Chulalongkorn Academic into Its 2nd Century Project Advancement Project (Thailand); the Welch Foundation, contract C-1845; and the Weston Havens Foundation (USA).

\end{acknowledgments}
\bibliography{auto_generated}

\cleardoublepage \appendix\section{The CMS Collaboration \label{app:collab}}\begin{sloppypar}\hyphenpenalty=5000\widowpenalty=500\clubpenalty=5000\vskip\cmsinstskip
\textbf{Yerevan Physics Institute,  Yerevan,  Armenia}\\*[0pt]
A.M.~Sirunyan,  A.~Tumasyan
\vskip\cmsinstskip
\textbf{Institut f\"{u}r Hochenergiephysik,  Wien,  Austria}\\*[0pt]
W.~Adam,  F.~Ambrogi,  E.~Asilar,  T.~Bergauer,  J.~Brandstetter,  E.~Brondolin,  M.~Dragicevic,  J.~Er\"{o},  A.~Escalante Del Valle,  M.~Flechl,  M.~Friedl,  R.~Fr\"{u}hwirth\cmsAuthorMark{1},  V.M.~Ghete,  J.~Grossmann,  J.~Hrubec,  M.~Jeitler\cmsAuthorMark{1},  A.~K\"{o}nig,  N.~Krammer,  I.~Kr\"{a}tschmer,  D.~Liko,  T.~Madlener,  I.~Mikulec,  E.~Pree,  N.~Rad,  H.~Rohringer,  J.~Schieck\cmsAuthorMark{1},  R.~Sch\"{o}fbeck,  M.~Spanring,  D.~Spitzbart,  W.~Waltenberger,  J.~Wittmann,  C.-E.~Wulz\cmsAuthorMark{1},  M.~Zarucki
% ...
\vskip\cmsinstskip
% ...
1:~Also at Vienna University of Technology,  Vienna,  Austria\\
% ...
\end{sloppypar}
\end{document}